\documentclass[12pt]{iopart}
\usepackage{graphicx}

\usepackage{float}

\usepackage{xcolor}
\usepackage{cite}
\usepackage{dsfont}
\usepackage{braket}
\usepackage{tikz}               

\newcommand{\mc}{\multicolumn{1}{c}}

\begin{document}
\title[]{Geometric potential for a Bose-Einstein condensate on a curved surface}

\author{Sheilla M. de Oliveira}
\address{Physics Department and Research Center OPTIMAS,
Rhineland-Palatinate Technical University Kaiserslautern-Landau, Erwin-Schr\"odinger Straße 46, 67663 Kaiserslautern, Germany, and Instituto de F\'isica de S\~ao Carlos, Universidade de S\~ao Paulo - 13566-590 S\~ao Carlos, SP, Brazil}

\author{Nat\'alia Salom\'e M\'oller}
\address{Research Center for Quantum Information, Institute of Physics, Slovak Academy of Sciences, D\'ubravsk\'a Cesta 9,
84511 Bratislava, Slovakia}
\ead{natalia.moller@savba.sk}
\date{\today}

\vspace{10pt}
\begin{indented}
	\item[]\today
\end{indented}

\begin{abstract}
We compute the ground state of a Bose-Einstein condensate confined on a curved surface and unravel the effects of curvatures. Starting with a general formulation for any smooth surface, we apply it to a prolate ellipsoid, which is inspired by recent bubble trap experiments. 
Using only elementary tools, we perform a perturbative approach to the Gross-Pitaevskii equation and a general Ansatz, followed by a dimensional reduction. We derive an effective two-dimensional equation that includes a curvature-dependent geometric potential.
We compute the ground state using Thomas-Fermi approximation and, for an isotropic confinement, we find that the highest accumulation of atoms happens on the regions with the greatest difference between the principal curvatures. For a prolate ellipsoid, this accumulation happens on the equator, which is contrary to previous findings that describe accumulation on the poles of a bubble trap. Finally, we explain the reasons for this difference: the higher accumulation of atoms on the poles happens due to anisotropies in the confinement, while the higher accumulation on the equator happens exclusively due to the geometric properties of the surface.  
\end{abstract}

\maketitle

\section{Introduction}

The physics of quantum systems constrained to curved manifolds of one and two-dimensions were already subject of interest in the last century, where effective low-dimensional Schr\"ondinger equations were derived~\cite{DeWitt, Jensen, daCosta}. Dimensional reduction techniques have found application in diverse fields, such as in electronic materials~\cite{Eletronic,Magnetism}, photonics~\cite{photonic}, DNA wires~\cite{DNA}, and more.
Of particular relevance to the present work, there is a variety of results of Bose-Einstein condensates (BEC) on curved wires and surfaces~\cite{ReviewTononi}.

Flat geometries are particular cases of curved geometries, where the curvatures vanish everywhere. Experiments of BECs on flat manifolds of one and two dimensions have already been performed by many groups with diverse techniques~\cite{ReviewBloch}, where cigar-shaped and disk-shaped traps are realized~\cite{Gorlitz,Rychtarik}. Curiously, it was initially believed to not be possible to implement a BEC on low dimensions, due to the Hohenberg-Mermin-Wagner theorem that predicts that the phase transition to a BEC happens only at null temperature in a homogeneous system~\cite{MerminWagner, 
Hohenberg}. Later, it was found that different boundary conditions for a low-dimensional flat trap allow the BEC formation at a positive temperature~\cite{BEC2Dsim}. Theoretical dimensional reduction techniques for these flat traps can be found, for instance, in references~\cite{Olshanii, JacksonAD, Chiofalo, LucaSalasnich}. 

The confinement of BECs on one-dimensional curved manifolds has already been experimentally demonstrated as well, where ring-shaped potentials have been quite often explored~\cite{Gupta, RyuC, Heathcote, Sherlock, Ramanathan, Moulder, RyuC2}. To cite a few applications, one can characterize the rotation of a ring~\cite{SchwartzStrigari}, its employment on atomtronics~\cite{Atomtronics}, the propagation of solitons \cite{Solitons1D} and sound waves~\cite{Sound1D}, and even the simulation of an early universe in expansion~\cite{Campbell}.
Ring-shaped traps are one specific kind of curved manifold, with the shape of a circular ring. One can also find the demonstration of more exotic shapes of one-dimensional traps~\cite{Any1d}. A theoretical analysis of a BEC on an ellipsoidal waveguide can be seen in~\cite{SchwartzStrigari, LucaSalasnich2} and on any kind of one-dimensional geometry in~\cite{BEC1D}.

Recently, the prospects of observing a Bose gas on a two-dimensional curved manifold became a reality with a type of confinement known as {\it bubble trap}~\cite{BubblesSpace, ExpJia}. This experiment was initially proposed in~\cite{OZobay0, OZobayPRL, OZobayPRA}, involving atoms dressed by a radio-frequency field trapped in an inhomogeneous magnetic field. With this, it is possible to let the gas be confined to the surface of an ellipsoidal shape. The first experimental realization of a Bose gas in thermal state on the surface of an ellipsoid using this technique were performed in~\cite{Colombe2004}. Attempts to reach the BEC state were performed in the same experiment~\cite{HelExp, GarHelReview, GarHelReview2}, however, it happened that the gas was found only on the bottom of this bubble due to the gravitational sag. Thus, it became necessary to find alternative routes to implement this bubble shape successfully.

The first one of these routes was to launch this experiment to the International Spacial Station (ISS) of NASA, where a cold atom laboratory was developed to perform a variety of experiments with fermionic and bosonic cold gases~\cite{Space1, Space2, Space3, Space4}. A little time after the first realization of a BEC in ISS~\cite{BECinSpace}, it was reported the first step on the bubble trap experiments: an ultra-cold Bose gas was confined on the surface of a bubble trap~\cite{BubblesSpace}, where the gas is found in thermal states. The second route to implement experimentally the bubble trap happened in a terrestrial laboratory using an alternative technique based on the use of dual atomic species~\cite{ExpJia}, a method proposed in~\cite{HoTL,PuH,WolfAntun}. Realizing the cold gas in the BEC state on a bubble trap is still being a challenge in all these experiments, and in~\cite{Perspectives, Perspectives2} one can find the perspectives for all these developments. Finally, further routes are also being explored to compensate gravity, and will enable the implementation a bubble trap with the radio-frequency dressing technique on the ground as well~\cite{ExpGuo}. 

Motivated by these experimental achievements, diverse topics concerning BECs on a bubble trap were explored theoretically. Collective modes were computed using diverse methods~\cite{BECmanifold, QuasiEllipsoid, 2007, KSun, KPadavic, Lima} which, in particular, can characterize the crossover from a full sphere to a hollowed one~\cite{KSun, KPadavic}. The critical temperature of a BEC on the surface of a sphere was computed in~\cite{Bereta, TononiSphere, Rhyno}, and the superfluid transition was characterized in~\cite{AndreaAxel}. The stability of the of the fluid phase was analyzed in~\cite{MonteCarlo, Ciardi} and of stationary states of a BEC in~\cite{Andriati, Brito}. The self-interference in a bubble trap in free expansion is studied in references~\cite{2007, LucaMonteCarlo}, and its thermodynamic properties in~\cite{Rhyno}. The authors of~\cite{MeisterRoura} propose a scheme that enables the generation of a slowly expanding shell with an isotropic momentum distribution. Furthermore, dipolar gases are explored on the surface of a sphere in~\cite{Dipole2012, Lima, Ciardi, GravitySag,Sanchez} and vortices dynamics in~\cite{BeretaPRA2022, PadavicVortices, Saito}, while atomic interaction properties are characterized in~\cite{Carmesin, Tonini2022}. A review of these topics can be seen in references~\cite{Perspectives2, TononiReview}

For simplicity, the majority of the theoretical analyses describe the bubble as a perfect sphere, even though an ellipsoidal shape would be a better approximation~\cite{GarHelReview}. Nevertheless, a sphere is a good approximation, including the fact that the topological properties of these two geometrical shapes are the same~\cite{AndreaAxel}. Among a few works that describe the bubble trap as an ellipsoid~\cite{TononiReview, Tononinew}, in reference~\cite{LucaMonteCarlo}, the authors compute the ground state of a BEC confined on such a surface. They find that the atoms show a higher accumulation on the poles of a prolate ellipsoid, and this agrees with the experimental realizations~\cite{BubblesSpace}.

In the present manuscript, we also compute the ground state of a BEC on an ellipsoid. In special, we show that the geometric properties of an ellipsoid lead to an opposite ground state profile as the one above described: they induce a higher accumulation of atoms on the equator of a prolate ellipsoid, and not on the poles. Here, we are going to derive these results and explain this difference.

We initially derive our results for any kind of curved surface, where we obtain a two-dimensional Gross-Pitaeviskii (2DGP) equation~(\ref{maingp}) and a Thomas-Fermi solution~(\ref{TFsolution}). Subsequently, we apply these equations to the specific case of an ellipsoid, allowing us to discuss applications on a bubble trap and make a comparison with the ground state profile from~\cite{BubblesSpace, LucaMonteCarlo}.
We work with a perturbative method, inspired by tools from reference~\cite{Lammerzahl}, followed by a dimensional reduction technique, where we integrate out the variable perpendicular to the surface. We obtain the 2DGP equation, where a term can be interpreted as a low-dimensional effective potential and depends on the curvatures of the surface. This is a term known as {\it geometric potential}~\cite{Jensen, daCosta}, and it is responsible for the accumulation of atoms on the equator of a prolate ellipsoid. 


We proceed as follows. In section~\ref{SecMath} we introduce the mathematical objects we are going to use, where many of these tools were already used in a previous manuscript from one of us~\cite{BECmanifold}. In section~\ref{SecModel} we establish the physical model we are going to work on: we define the hypotheses on the confinement frequency and on the strength of the interactions. In section~\ref{PertExp} we develop the perturbative expansion, in section~\ref{SecDimRed} we employ the dimensional reduction and in section~\ref{SecTF} we explore the Thomas-Fermi solution, for any kind of surface. In section~\ref{SecEllipsoid} we define an ellipsoid inspired by the experimental parameters of a bubble trap and in section~\ref{SecBubble} we compare our results with the literature. In section~\ref{SecDisc} we present a discussion and in section~\ref{SecConc} our conclusions.



\section{Mathematical preliminaries} \label{SecMath}
In this section, we introduce the mathematical objects that are necessary for the development of the paper. We closely follow the methods of a previous work by one of us~\cite{BECmanifold}, where the reader can find a similar introductory section. The content of this section can be found in textbooks~\cite{doCarmo, Lee, Jost} and other references, such as~\cite{GaussCoord, GaussCoordSyst, 3form}.

Let us consider a smooth surface $\mathcal{M}$, that is, a two-dimensional differentiable manifold embedded in the three-dimensional Euclidean space. Assume that a given portion of $\mathcal{M}$ can be described by the two variables $x^1$ and $x^2$. That is, any point ${\bf p}$ that belongs to this surface portion can be expressed as ${\bf p}=(x^1,x^2)$. Since $\mathcal{M}$ belongs to the 3D space, these two variables might be expressed as functions of the Cartesian coordinates $(x,y,z)$.
On a surface, it is possible to define curvatures in more than one way. Important definitions are the principal curvatures $\kappa_1$ and $\kappa_2$, the mean curvature $H$, and the Gaussian curvature $K$.

To define the principal curvatures, let us first define the curvature of a curve on a plane, that is, of a one-dimensional manifold embedded in the two-dimensional Euclidean space. A curve on a plane can be described by a parametrized equation of the type $\gamma(t)=(x(t),y(t))$, such that $\|\gamma'(t)\|=\sqrt{x'(t)^2+y'(t)^2}=1$. The tangent vector to this curve is given by $\gamma'(t)$, and its curvature is given by $\kappa(t)=\|\gamma''(t)\|$. The interpretation of this curvature is that it is the inverse of the radius of the circle that is the best approximation to the curve $\gamma(t)$ at the respective point.

Now, let us consider again the surface $\mathcal{M}$, a specific point ${\bf p}(x^1,x^2)$ on this surface and the normal vector ${\bf n}({\bf p})$ to $\mathcal{M}$ at ${\bf p}$. There is an infinite set of planes with origin at ${\bf p}$ and parallel to ${\bf n}({\bf p})$. Each of these planes intersects the manifold $\mathcal{M}$, and this intersection is a curve on a plane, as described in the previous paragraph. Take this infinite set of curves and their respective curvatures at ${\bf p}$, and consider the maximum and minimum among these curvatures. This defines the principal curvatures $\kappa_1(x^1,x^2)$ and $\kappa_2(x^1,x^2)$.

The mean and Gaussian curvatures are defined as
\begin{equation}
\label{meanCurv}
H(x^1,x^2)=\frac{\kappa_1(x^1,x^2)+\kappa_2(x^1,x^2)}{2}
\end{equation}
and
\begin{equation}
\label{GaussianCurv}
K(x^1,x^2)=\kappa_1(x^1,x^2)\kappa_2(x^1,x^2).
\end{equation}
Conversely, one can also obtain that
\begin{equation}
\label{kappaHK}
\kappa_i(x^1,x^2)=H(x^1,x^2)\pm\sqrt{H^2(x^1,x^2)-K(x^1,x^2) \ }.
\end{equation}

Remember that the mean curvature of a curve at a given point is the inverse of the radius of the circle that best fits on a curve at this point. For the surface $\mathcal{M}$, let us define
\begin{equation}
R(x^1,x^2)=\min
  \left\{
    \frac{1}{\kappa_1(x^1,x^2)},
    \frac{1}{\kappa_2(x^1,x^2)}
  \right\},
\end{equation}
which is the radius of the smaller circle that fits on a path on the surface at ${\bf p}=(x^1,x^2)$, as we explained in the above paragraph. It is also possible to define the minimum among the radii $R(x^1,x^2)$:
\begin{equation}
R=\min\{R(x^1,x^2), \ \mathrm{for} \ (x^1,x^2)\in\mathcal{M}\}.
\end{equation}
For a compact surface, $R$ is strictly positive.

Let us now consider the three-dimensional space in the vicinity of a portion of the smooth surface $\mathcal{M}$ that contains ${\bf p}=(x^1, x^2)$. Consider a point ${\bf q}$ in the vicinity of $\mathcal{M}$ and suppose that the distance $d\{ {\bf q},\mathcal{M}\}$ between  ${\bf q}$ and $\mathcal{M}$ is given by the distance $d\{ {\bf q},{\bf p}\}$ between ${\bf q}$ and ${\bf p}$. That is, ${\bf p}$ is the point on $\mathcal{M}$ that is the closest to ${\bf q}$.

With this, we can introduce the Gaussian normal coordinate system that describes the three-dimensional space and is represented by the variables $(x^0,x^1,x^2)$. The variable $x^0$ is defined as the distance from a given point to the manifold, and the variables $(x^1,x^2)$ are defined as their values on the point on $\mathcal{M}$ that is the closest. That is, in the Gaussian normal coordinate system, the point ${\bf q}$ as described above is given by
\begin{equation}
{\bf q}=(x^0,x^1,x^2)
=(d\{ {\bf q},{\bf p}\}, x^1, x^2),
\end{equation}
while the point ${\bf p}$ that is on the surface is given by
\begin{equation}
{\bf p}
=(0, x^1, x^2).
\end{equation}
The Gaussian normal coordinate system is well-defined only in the vicinity of the manifold, where $|x^0|<R$. With this, each point is described by a unique combination of variables $x^0$, $x^1$, and $x^2$. For simplicity, we will consider $-R/2< x^0< R/2$. With this, we define the neighbourhood $\mathcal{N}(\mathcal{M})$ of $\mathcal{M}$ as the points ${\bf q}=(x^0,x^1,x^2)$ of
the 3D space described by the Gaussian normal coordinate system, with $|x^0| < R/2$. 

The metric of a manifold is defined as a matrix whose entrances are the internal product between the tangent vectors:
\begin{equation}
G_{\mu\nu}={\bf v}_\mu\cdot {\bf v}_\nu,
\end{equation}
where ${\bf v}_\mu=\partial {\bf q}/\partial x^\mu$, for $\mu=0,1,2$. The metric of the three-dimensional space in the Gaussian normal coordinates in the vicinity of $\mathcal{M}$ is given by~\cite{BECmanifold}
\begin{equation}G_{\mu\nu}(x^0,x^1,x^2)= 
\renewcommand{\arraystretch}{1.2}
  \left(
  \begin{array}{ c | c c }
    \mc{1} & \ \ \ \ \ \ 0 & \ 0 \\
    \cline{2-3}
    0  & &  \\
    0  & \multicolumn{2}{c}{\raisebox{.6\normalbaselineskip}[0pt][0pt]{$g_{ij}(x^0,x^1,x^2)$}} \\
  \end{array}
  \right), \label{metric}
\end{equation}
and the metric of the surface $\mathcal{M}$ is given by 
\begin{equation}\label{g_0}
(g_0)_{ij}(x^1,x^2)
=
g_{ij}(0,x^1,x^2).
\end{equation}

From Appendix A of reference~\cite{BECmanifold}, one can see the relation of the metric in the Gaussian normal coordinate system and the above curvatures:
\begin{equation}\label{detgx0}
\hspace{-1.5cm}\det g(x^0)=\det g_0
\left[
 1+4x^0 H
 +(x^0)^2(4H^2+2K)
 +4(x^0)^3HK
 +(x^0)^4K^2
\right],
\end{equation}
where we hide the dependence of $g$, $g_0$, $H$ and $K$ on $x^1$ and $x^2$ for brevity. Note that this is an exact formula. Moreover, this metric determinant can be re-expressed with the principal curvatures as
\begin{equation}
\hspace{-1.5cm}\det g(x^0)=\det g_0 (1+\kappa_1x^0)^2(1+\kappa_2x^0)^2. \label{detgbom}
\end{equation}

The Laplacian in Gaussian normal coordinates can be written as
\begin{equation}
\Delta=\frac{\partial^2}{\partial {x^0}^2}+
\frac{\partial}{\partial x^0}\left(\ln\sqrt{\det g(x^0) \ }\right) \frac{\partial}{\partial x^0}+\Delta_{\mathcal{M}(x^0)},\label{LaplacCurv1}
\end{equation}
with the abbreviation
\begin{equation}
\Delta_{\mathcal{M}(x^0)}=\frac{1}{\sqrt{\det g(x^0) \ }}\frac{\partial}{\partial x^i}\left(\sqrt{\det g(x^0) \ } g^{ij}(x^0)\frac{\partial}{\partial x^j}
\right). \label{DeltaM}
\end{equation}

All the mathematical objects reviewed here are going to be used in the following sections within a physical model. 
 
\section{Model}\label{SecModel}

To study a Bose gas that is confined on a curved surface we must initially establish how it is confined there. For that, we are going to consider a modified harmonic trap as the following. Significantly, this confinement must be strong, so that the gas behaviour becomes almost two-dimensional.

To that, let us choose a harmonic frequency $\omega_0$, with the condition that the corresponding one-dimensional cloud width 
$\sigma_0=\sqrt{\hbar/m\omega_0}$ satisfies $\sigma_0/R\ll 1$. Here, $\hbar$ is the Planck constant and $m$ is the mass of the confined particles. 

We consider then the following confinement potential:
\begin{eqnarray}
\label{Utrap}
\hspace{-1cm}
U_\mathrm{trap}(x^0,x^1,x^2)&=
\frac{m[\omega_0+\omega_2(x^1,x^2)]^2}{2}(x^0)^2
\hspace{4.15cm}
\\
&=
\frac{m\omega_0^2}{2}(x^0)^2
+m\omega_0\omega_2(x^1,x^2)(x^0)^2
+\frac{m\omega_2(x^1,x^2)^2}{2}(x^0)^2.
\nonumber
\end{eqnarray}
By definition, we have that
\begin{equation}
\label{orderomegao}
\mathcal{O}\left(\frac{\omega_0}{\omega_R}\right)=\mathcal{O}\left(\left[\frac{\sigma_0}{R}\right]^{-2}\right),
\end{equation}
where we introduced $\omega_R=\hbar/mR^2$ for a dimensionless comparison of orders of magnitude. Importantly, we impose that
\begin{equation}
\label{orderomega2}
\mathcal{O}\left(\frac{\omega_2(x^1,x^2)}{\omega_R}\right)=\mathcal{O}\left(1\right),
\end{equation}
for all $x^1$ and $x^2$ where the Gaussian normal coordinate system is well-defined. 

Note that, for completeness, we could have added a term $m\omega_1(x^0)^2/2$, with $\mathcal{O}(\omega_1/\omega_R)=\mathcal{O}\left((\sigma_0/R)^{-1}\right)$ in the above potential. However, we would have imposed it to be constant as well, thus it can be absorbed into the factor $\omega_0$ without any loss of generality. The reason we impose these frequencies to be constants is that, without these constraints, it is not possible to perform a dimensional reduction for any infinitely thin manifold. In general, divergent terms would appear in the limit $\sigma_0\rightarrow 0$, and the gas would accumulate at a specific point rather than being distributed across the surface. A well-defined dimensional reduction was performed for a case where $\omega_0$ is not constant in~\cite{QuasiEllipsoid}, where the studied surface is asymptotically a sphere.

The wave function $\Psi(x^0,x^1,x^2)$ that is a solution or an approximate solution for the ground state of a gas confined in the potential (\ref{Utrap}) for $\sigma_0\ll R$ is expected to be concentrated in the neighbourhood of the surface $\mathcal{M}$. That is, we expect that a good approximation for the particle number is given by
\begin{equation}
N\simeq\int_{-R/2}^{R/2} dx^0 \int dx^1 dx^2 \sqrt{g(x^0,x^1,x^2)}\|\Psi(x^0,x^1,x^2)\|^2.
\end{equation}
Moreover, any other quantity related to the wave function can also be computed considering only this spatial region. 

Finally, we consider that the interaction strength $g_\mathrm{int}$ among the atoms is weak. Specifically, we consider that 
\begin{equation}
\mathcal{O}\left(\frac{a_s}{R}\right)=\mathcal{O}\left(\frac{\sigma_0}{R}\right),
\end{equation}
where $a_s$ is the scattering length with the relation
$g_\mathrm{int}=4\pi\hbar^2a_s/M$. This hypothesis is not strictly necessary to study a dimensional reduction. For instance, references~\cite{KPadavic, KSun} study the behavior of a quasi-two-dimensional sphere considering strong interactions. Here we make this choice of such weak interactions in order to study the influence of interactions just after performing a dimensional reduction.

\section{Perturbative expansion} \label{PertExp}

In this section, we develop part of our methodical calculations. Readers who prefer to skip the technical details may read only the next paragraph, where we summarize the main steps of this section, and the last two paragraphs, where we motivate the calculations of section~\ref{SecDimRed}.

Here, we consider a general Ansatz for the wave function given by equation~(\ref{ansatz0}), with the conditions~(\ref{OrdersS0}) and (\ref{OrdersS1}). We insert this Ansatz into the GP equation, which is given by expression~(\ref{GP}), in a short form, and by expression~(\ref{GPlong1}), in an extended form. We follow a perturbative expansion in orders $\mathcal{O}(\sigma_0/R)$. The first order to be considered is $\mathcal{O}([\sigma_0/R]^{-2})$, from which we obtain the solution for the term $e^{S_0}$ from the Ansatz~(\ref{ansatz0}), which is proportional to the Gaussian wave function $\mathcal{G}$ from equation~(\ref{solutionHarmOsc}). This term guarantees that the wave function is exponentially small far from the surface~$\mathcal{M}$. We proceed to the next order $\mathcal{O}([\sigma_0/R]^{-1})$. From this step, we obtain a partial solution for the term $e^{S_1}$ from the Ansatz~(\ref{ansatz0}), which is given by equation~(\ref{es1}). From these findings, we can express the Ansatz as in equation (\ref{finalAnsatz}), where $\psi$ must satisfy equation~(\ref{GP3}) up to order $\mathcal{O}(1)$. Equation~(\ref{GP3}) thus is the development of the GP equation~(\ref{GP}) with the imposed Ansatz.

With~(\ref{finalAnsatz}) and (\ref{GP3}), it is possible to proceed to section~\ref{SecDimRed}. In the following, we perform the above-described systematic calculations. 

We intend to solve approximately the Gross-Pitaevskii (GP) equation
\begin{eqnarray}
-\mu\Psi-\frac{\hbar^2}{2m}\Delta\Psi+U_\mathrm{trap}
+g_{\rm int}|\Psi|^2\Psi=0. \label{GP}
\end{eqnarray}
which can be denoted as
\begin{equation}
\mathrm{GP}(\Psi)=0.
\end{equation}
Inserting the potential~(\ref{Utrap}) and the expression~(\ref{LaplacCurv1}) for the Laplacian into the above equation, the GP equation in Gaussian normal coordinates becomes
\begin{eqnarray}
\label{GPlong1}
-\mu\Psi-\frac{\hbar^2}{2m}
\left(
  \frac{\partial^2}{\partial (x^0)^2}
  +
  \left[
    \frac{\partial \ln\sqrt{\det g}}{\partial x^0}
  \right]
  \frac{\partial}{\partial x^0}
  +\Delta_{\mathcal{M}(x^0)}
\right)\Psi
\\
\nonumber
+\frac{m\omega_0^2}{2}(x^0)^2\Psi
+m\omega_0\omega_2(x^0)^2\Psi
+\frac{m\omega_2^2}{2}(x^0)^2\Psi
+g_{\rm int}|\Psi|^2\Psi=0. 
\end{eqnarray}
To solve approximately this equation, we are going to consider the following Ansatz:
\begin{equation}\label{ansatz0}
\Psi(x^0,x^1,x^2)=\alpha^{1/2} e^{S_0(x^0,x^1,x^2)+S_1(x^0,x^1,x^2)+...} \ ,
\end{equation}
where $\alpha$ is a constant defined by the units chosen to describe the problem. Importantly, close to the surface $\mathcal{M}$, specifically, where $\mathcal{O}(x^0/R)=\mathcal{O}(\sigma_0/R)$, we assume the following:
\begin{eqnarray}
\label{OrdersS0}
\mathcal{O}
  \left(
    R\frac{\partial S_0}{\partial x^0}
  \right)
=
\mathcal{O}\left(\left[\frac{\sigma_0}{R}\right]^{-1}\right)
\ , \ \ \ \ 
\mathcal{O}
  \left(
    R^2\frac{\partial^2 S_0}{\partial (x^0)^2}
  \right)
=
\mathcal{O}\left(\left[\frac{\sigma_0}{R}\right]^{-2}\right);
\\
\nonumber
\\
\mathcal{O}
  \left(
    R\frac{\partial S_1}{\partial x^0}
  \right)
=
\mathcal{O}
  \left(
    R^2\frac{\partial^2 S_1}{\partial (x^0)^2}
  \right)
=
\mathcal{O}(1).
\hspace{3.9cm}
\label{OrdersS1}
\end{eqnarray}
The above hypotheses (\ref{OrdersS0}) and (\ref{OrdersS1}) are motivated by the assumptions~(\ref{orderomegao}) and (\ref{orderomega2}), and inspired by the approach developed in reference~\cite{Lammerzahl}. Note that $S_0$ represents the term responsible for $\Psi$ being very sharp in the vicinity of the surface $\mathcal{M}$.

Generally, we also consider that 
$
\mathcal{O}
  \left(
    R\partial S_n/\partial x^0
  \right)
=
\mathcal{O}([\sigma_0/R]^{n-1})
$
and 
$
\mathcal{O}
  \left(
    R^2\partial^2 S_n/\partial (x^0)^2
  \right)
=
\mathcal{O}([\sigma_0/R]^{n-2})
$, for $x^0\simeq \sigma_0$ and $n\geq 2$. However, we will need only terms until order $\mathcal{O}(1)$ to reach our results, meaning that we will neglect terms of orders $\mathcal{O}(\sigma_0/R)$ or higher. With this only the terms $S_0$ and $S_1$ are relevant to our calculations. The terms associated with $S_2$ vanish under the approximations we are making, and we show it in \ref{AppS2}. To be precise, we would need to include the terms associated with $S_2$ in the following calculations, but for simplicity, we will already now ignore them. Moreover, the terms associated with $S_n$, for $n\geq 3$ are of order $\mathcal{O}(\sigma_0/R)$ or higher, and we already neglect them now.

Evaluating the first and second derivatives of $\Psi$, we obtain
\begin{equation}
\hspace{-1cm}
\frac{\partial \Psi}{\partial x^0}
=
(S'_0+S'_1)\Psi
\ \ \ \ \mathrm{and} \ \ \ \ 
\frac{\partial^2 \Psi}{\partial (x^0)^2}
=
(S''_0+{S'_0}^2+2S'_0S'_1+S''_1+{S'_1}^2)\Psi.
\end{equation}
Inserting these expressions with the above Ansatz into the GP equation and dividing it by $\Psi$, we obtain approximately that
\begin{eqnarray}
\hspace{-2.3cm}
-(\mu_0+\mu_1+\mu_2)
-\frac{\hbar^2}{2m}
\left(
  S''_0+{S'_0}^2+2S'_0S'_1+S''_1+{S'_1}^2
  +
  \left[
    \frac{\partial \ln\sqrt{\det g}}{\partial x^0}
  \right]
  (S'_0+S'_1)
\right)
\nonumber
\\
\nonumber
\\
\hspace{-1cm}
-\frac{\hbar^2}{2m}\frac{\Delta_{\mathcal{M}(x^0)}\Psi}{\Psi}
+\frac{m\omega_0^2}{2}(x^0)^2
+m\omega_0\omega_2(x^0)^2
+\frac{m\omega_2^2}{2}(x^0)^2
+g_{\rm int}|\Psi|^2=0.
\label{GP1}
\end{eqnarray}
where we divided the chemical potential into three terms of orders
\begin{equation}
\hspace{-1cm}
\mathcal{O}
\left(
  \frac{\mu_0}{\mu_R}
\right)
=
\mathcal{O}{
\left(
  \left[
    \frac{\sigma_0}{R}
  \right]^{-2}
\right)}
, \ \ \ 
\mathcal{O}
\left(
  \frac{\mu_1}{\mu_R}
\right)
=
\mathcal{O}{
\left(
  \left[
    \frac{\sigma_0}{R}
  \right]^{-1}
\right)}
, \ \ \
\mathcal{O}
\left(
  \frac{\mu_2}{\mu_R}
\right)
=
\mathcal{O}{(1)},
\end{equation}
where $\mu_R=\hbar\omega_R/2$ is introduced for a dimensionless comparison.

Notably, the above equation is valid only for the range where
\begin{equation}
\label{rangex0}
\mathcal{O}\left(\frac{x^0}{R}\right)=\mathcal{O}\left(\frac{\sigma_0}{R}\right).
\end{equation}
This constraint occurs because we already have that $\Psi\simeq 0$ outside this range, and consequently, the GP equation~(\ref{GPlong1}) is approximately satisfied automatically. Therefore, we can only impose equation~(\ref{GP1}) to be approximately valid in the range~(\ref{rangex0}).

Now, considering only the terms of order $\mathcal{O}([\sigma_0/R]^{-2})$ in equation~(\ref{GP1}), we obtain
\begin{eqnarray}
-\mu_0
-\frac{\hbar^2}{2m}
\left(
  S''_0+{S'_0}^2
\right)
+\frac{m\omega_0^2}{2}(x^0)^2
=0.
\label{checkzero1}
\end{eqnarray}
where we remind that $\mathcal{O}(\omega_0^2/\omega_R^2)=\mathcal{O}([\sigma_0/R]^{-4})$ and $\mathcal{O}([x^0]^2/R^2)=\mathcal{O}(\sigma_0^2/R^2)$, leading to $\mathcal{O}(\omega_0^2[x^0]^2/[\omega_R^2R^2])=
\mathcal{O}([\sigma_0/R]^{-2})$.

Denoting $\mathcal{G}=\pi^{-1/4}\sigma_0^{-1/2}e^{S_0}$, we have that $\mathcal{G}''=( S''_0+{S'_0}^2)\mathcal{G}$ and the above equation is simply the equilibrium equation of a one-dimensional quantum harmonic oscillator:
 \begin{eqnarray}
-\mu_0\mathcal{G}
-\frac{\hbar^2}{2m}
\mathcal{G}''
+\frac{m\omega_0^2}{2}(x^0)^2\mathcal{G}
=0,
\end{eqnarray}
leading to the following Gaussian wave-function and chemical potential
\begin{equation}
\label{solutionHarmOsc}
\mathcal{G}(x^0)=\frac{e^{-(x^0)^2/2\sigma_0^2}}{\sqrt[4]{\pi}\sqrt{\sigma_0}}
\ , \ \ \ \
\mu_0=\frac{\hbar\omega_0}{2}.
\end{equation}
Consequently, we obtain that
\begin{equation} \label{S0'}
S_0(x^0)=-\frac{(x^0)^2}{2\sigma_0^2}
\ , \ \ \ \
S'_0(x^0)=-\frac{x^0}{\sigma_0^2}
\ , \ \ \ \
S''_0(x^0)=-\frac{1}{\sigma_0^2}.
\end{equation}
Note that this solution is consistent with hypotheses~(\ref{OrdersS0}) and that $S_0$ is independent on the variables $x^1$ and $x^2$. With this, the ansatz~(\ref{ansatz0}) can be rewritten as
\begin{equation}
\label{ansatz}
\Psi(x^0,x^1,x^2)
=
\mathcal{G}(x^0)
\psi(x^0,x^1,x^2),
\end{equation}
where $\psi(x^0,x^1,x^2)=\pi^{1/4}\sigma_0^{1/2}\alpha^{1/2}e^{S_1(x^0,x^1,x^2)}$. Remember that here we are already ignoring the terms $S_n$, for $n\geq 2$.

With the expression~(\ref{ansatz}), the GP equation~(\ref{GP1}) becomes
\begin{eqnarray}
-(\mu_1+\mu_2)
-\frac{\hbar^2}{2m}
\left(
  2S'_0S'_1+S''_1+{S'_1}^2
  +
  \left[
    \frac{\partial \ln\sqrt{\det g}}{\partial x^0}
  \right]
  (S'_0+S'_1)
\right)
\nonumber
\\
\nonumber
\\
-\frac{\hbar^2}{2m}\frac{\Delta_{\mathcal{M}(x^0)}\psi}{\psi}
+m\omega_0\omega_2(x^0)^2
+\frac{m\omega_2^2}{2}(x^0)^2
+g_{\rm int}\mathcal{G}^2|\psi|^2=0,
\label{GP2}
\end{eqnarray}
which is also valid only in the regime~(\ref{rangex0}).
Now, considering only the terms of order $\mathcal{O}([\sigma_0/R]^{-1})$, the above equation becomes:
\begin{eqnarray}
-\mu_1+\frac{\hbar^2 x^0}{m\sigma_0^2}
\left(
  S'_1
  +\frac{1}{2}
  \left[
    \frac{\partial \ln\sqrt{\det g}}{\partial x^0}
  \right]
\right)
\label{checkzero2}
=0.
\end{eqnarray}
which imposes that $\mu_1=0$ (see \ref{Appmu1}) and that
\begin{eqnarray}
\frac{\partial S_1}{\partial x^0}
=\frac{\partial \ln\left( (\det g)^{-1/4}\right)}{\partial x^0}
\label{solutionS1}
\end{eqnarray}
Integrating this equation on $x^0$ we obtain
\begin{eqnarray}
S_1(x^0,x^1,x^2)
=\ln\left( [\det g(x^0,x^1,x^2)]^{-1/4}\right)+\bar{S_1}(x^1,x^2),
\end{eqnarray}
and then
\begin{equation}
\label{es1}
e^{S_1(x^0,x^1,x^2)}
=
\frac{e^{\bar{S}_1(x^1,x^2)}}{\sqrt[4]{\det g(x^0,x^1,x^2)}}
\end{equation}
Denote $\phi(x^1,x^2)=\pi^{1/4}(\alpha\sigma_0)^{1/2}e^{\bar{S}_1(x^1,x^2)}/\sqrt[4]{\det g_0(x^1,x^2)}$. Considering the expression~(\ref{solutionHarmOsc}), the Ansatz~(\ref{ansatz0}) takes the form
\begin{equation}
\hspace{-1cm}
\Psi(x^0,x^1,x^2)
=
\mathcal{G}(x^0)
\frac{\sqrt[4]{\det g_0(x^1,x^2)}}{\sqrt[4]{\det g(x^0,x^1,x^2)}}
\phi(x^1,x^2)
\label{Ansatz1}
\end{equation}
which is the same Ansatz as the proposed one in the reference~\cite{BECmanifold}, but written in a slightly different way. There, the term ${\sqrt[4]{\det g(x^0,x^1,x^2)}}$ was introduced to maintain the error in the calculation of the particle number exponentially small. Without this term, this error would have a quadratic decrease within the ratio $\sigma_0/R$. In the present paper, we derived this Ansatz from more general hypotheses and more elementary methods. Moreover, the dimensional reduction in reference~\cite{BECmanifold} was treated in a quasi-two-dimensional limit so that the dependence of $\sigma_0$ on the variables $x^1$ and $x^2$ was considered. In the present paper, we consider $\sigma_ 0$ as a constant.

From equation~(\ref{detgbom}), the Ansatz~(\ref{Ansatz1}) can be rewritten as
\begin{equation}
\label{finalAnsatz}
\Psi(x^0,x^1,x^2)
=
\mathcal{G}(x^0)
\psi(x^0,x^1,x^2)
=
\frac{\mathcal{G}(x^0)\phi(x^1,x^2)}{\sqrt{(1+\kappa_1x^0)(1+\kappa_2x^0)}} 
\end{equation}
Remember that this solution is written in Gaussian normal coordinate systems, which is well-defined only for $|x^0|< R$, where $R=\min\{1/\kappa_1,1/\kappa_2\}$. Consequently, the term $\sqrt{(1+\kappa_1x^0)(1+\kappa_2x^0)}$ never vanishes.

Turning back our attention to the GP equation, from the solution~(\ref{solutionS1}), we obtain that expression~(\ref{GP2}) becomes
\begin{eqnarray}
\label{GP5}
\hspace{-1cm}
-\mu_2\mathcal{G}\psi
-\frac{\hbar^2}{2m}
\left(
  -\frac{1}{2}
  \left[
    \frac{\partial^2 \ln\sqrt{\det g}}{\partial (x^0)^2}
  \right]
  -\frac{1}{4}
  \left[
    \frac{\partial \ln\sqrt{\det g}}{\partial x^0}
  \right]^2
\right)\mathcal{G}\psi
\\
\hspace{-1cm}
\nonumber
-\frac{\hbar^2}{2m}(\Delta_{\mathcal{M}(x^0)}\psi)\mathcal{G}
+m\omega_0\omega_2(x^0)^2\mathcal{G}\psi
+\frac{m\omega_2^2}{2}(x^0)^2\mathcal{G}\psi
+g_{\rm int}\mathcal{G}^3|\psi|^2\psi=0.
\end{eqnarray}
Here, note that $\Psi$ does not appear in the denominator anymore. Thus, the above equation is valid for all $x^0$ where the Gaussian normal coordinate system is well defined, that is, for $-R/2<x^0<R/2$.

Moreover, observe that equations (\ref{checkzero1}) and (\ref{checkzero2}) exactly vanish. This can be verified by checking these equations with the Ansatz~(\ref{ansatz}).

Reminding that $m\omega_0=\hbar/\sigma_0^2$ and using equation~(\ref{detgbom}), the above expression~(\ref{GP5}) takes the form
\begin{eqnarray}
\label{GP3}
-\mu_2\mathcal{G}\psi
-\frac{\hbar^2}{8m}
\left(
  \frac{\kappa_1}{1+\kappa_1x^0}
  -
  \frac{\kappa_2}{1+\kappa_2x^0}
\right)^2\mathcal{G}\psi
-\frac{\hbar^2}{2m}(\Delta_{\mathcal{M}(x^0)}\psi)\mathcal{G}
\\
\ \nonumber
\\
+\frac{\hbar\omega_2}{\sigma_0^2}(x^0)^2\mathcal{G}\psi
+\frac{m\omega_2^2}{2}(x^0)^2\mathcal{G}\psi
+g_{\rm int}\mathcal{G}^3|\psi|^2\psi=0.
\end{eqnarray}
To further evaluate this expression, we would need to compute higher orders of the ansatz $\Psi$, taking into account the terms $S_2$, $S_3$, and so on. However, we decided to only consider terms until order $\mathcal{O}(1)$. Thus, we can neglect $S_n$, for $n\geq 2$, and need to find an approximate solution for $\psi(x^0,x^1,x^2)$. To do that, remember that we assume $\sigma_0\ll R$, meaning that the function $\mathcal{G}(x^0)$ is extremely sharp. This means that the value of $\psi(x^0,x^1,x^2)$ is relevant only for small values of $x^0$, which is where $\psi(x^0,x^1,x^2)\simeq \phi(x^1,x^2)$.
Motivated by this reasoning, we will make a dimensional reduction, which allows us to define a two-dimensional Gross-Pitaevski (2DGP) equation for the two-dimensional wave function $\phi(x^1,x^2)$. 

A second reason for choosing to perform a dimensional reduction is that, differently from the previous steps, where we computed the terms of orders $\mathcal{O}([\sigma_0/R]^{-2})$ and $\mathcal{O}([\sigma_0/R]^{-1})$, we now have to consider the nonlinear term $g_{\rm int}\mathcal{G}^3|\psi|^2\psi$. In the previous steps, we factorized the term $\mathcal{G}(x^0)$ and played with polynomials, where we could clearly identify the orders of magnitude in $\sigma_0/R$ for all terms, considering that $\mathcal{O}(x^0/R)=\mathcal{O}(\sigma_0/R)$. Now, it is not clear how to evaluate the orders of magnitude with the functions $\mathcal{G}(x^0)$ and $\mathcal{G}^3(x^0)$, since we cannot factorize these terms anymore. With the dimensional reduction method, we will recover the transparency on the orders of magnitude.

\section{Dimensional reduction}  \label{SecDimRed}

In the previous section, we derived an approximate Anzats for the wave function of a gas on a surface $\mathcal{M}$ with the confinement potential given by~(\ref{Utrap})--(\ref{orderomega2}). The wave function is given by expression (\ref{finalAnsatz}), where the Gaussian wave function $\mathcal{G}$ is found in expression~(\ref{solutionHarmOsc}), and the term $\psi(x^0,x^1,x^2)=\phi(x^1,x^2)/\sqrt{(1+\kappa_1x^0)(1+\kappa_2x^0)}$ must satisfy equation~(\ref{GP3}) up to order $\mathcal{O}(1)$. In this section, we are going to perform a dimensional reduction on equation~(\ref{GP3}), leading to a 2DGP equation that is satisfied by the two-dimensional wave function $\phi(x^1,x^2)$. 

In this procedure, we multiply equation equation~(\ref{GP3}) by the function $\sqrt{(1+\kappa_1x^0)(1+\kappa_2x^0)} \ \mathcal{G}$ and integrate it only with respect to the $x^0$ variable on the limit $\sigma_0\rightarrow 0$. The reasons for the multiplicative terms are computational convenience and for guaranteeing well-defined integrals in the limit of a thin shell, which are thoroughly explained in the following. Readers who prefer to skip the technical details of this section may consider simply equations~(\ref{g2d}), (\ref{Vgeom})--(\ref{maingp}). 

To derive the dimensional reduction of equation~(\ref{GP3}), observe that its left side vanishes up to order $\mathcal{O}(1)$. We can multiply this equation by any function that admits a Taylor expansion of the type $1+\sum_n a_n (x^0)^n$, with $\mathcal{O}(a_n)=\mathcal{O}(R^{-n})$, without altering its leading-order terms. Note that the expression $\sqrt{(1+\kappa_1 x^0)(1+\kappa_2 x^0)}$ features exactly this kind of Taylor expansion. Therefore, for computational convenience, we multiply equation~(\ref{GP3}) by this expression.

Moreover, it is expected that the integral of the left side of~(\ref{GP3}) would vanish, since the integral of zero should remain zero. However, this is a subtle process in the limit of a thin shell, where $\omega_0\rightarrow\infty$ and $\sigma_0\rightarrow 0$. The Gaussian distribution, which is the square of the Gaussian wave function, is a well-defined function for finite $\sigma_0$ and it tends to the Dirac delta distribution in the thin shell limit: $\lim_{\sigma_0\rightarrow 0}\mathcal{G}^2(x^0)= \delta(x^0)$. According to the theory of distributions \cite{distributions}, the Dirac delta $\delta(x^0)$ is a well-defined mathematical object. The limit of the following integral can be computed as
\begin{equation}
\label{rep1delta}
\lim_{\sigma_0\rightarrow 0} \int_{\mathcal{N}\{x^0=0\} } f(x^0) \mathcal{G}^2(x^0) dx^0
=
\int_{\mathcal{N}\{x^0=0\} } f(x^0) \delta(x^0) dx^0 = f(0)
\end{equation}
where the integral is performed on any neighbourhood $\mathcal{N}\{x^0=0\}$ around $x^0=0$, and $f(x^0)$ is any well-behaved function on this neighbourhood. For simplicity, let us consider that this neighbourhood is given by the interval $[-R/2,R/2]$, in which we can guarantee that the Gaussian normal coordinate system is well defined.

However, note that the limit $\lim_{\sigma_0\rightarrow 0}\mathcal{G}(x^0)$ is not a well-defined mathematical object. As a result, simply integrating the left side of equation~(\ref{GP3}) with respect to $x^0$ in the limit of a thin shell would not constitute a valid mathematical operation. To perform a proper integration, we must multiply equation~(\ref{GP3}) by $\mathcal{G}$, and afterwards integrate it with respect to $x^0$. 

This argument can be applied directly to nearly all the terms of equation~(\ref{GP3}), including the term $(\hbar\omega_2/\sigma_0^2)(x^0)^2\mathcal{G}\psi$, even with the dependence of its quotient $(x^0)^2/\sigma_0^2$ on $\sigma_0$. Although this may not be immediately evident, the limit below is a well-defined distribution
\begin{equation}
\label{rep2delta}
\lim_{\sigma_0\rightarrow 0}\frac{(x^0)^2}{\sigma_0^2}\mathcal{G}^2(x^0)=
\frac{\delta(x^0)}{2}
\end{equation}
as one can check in~\ref{AppInteg}.

The non-linear term $g_{\rm int}\mathcal{G}^3|\psi|^2\psi$ is the only one we cannot directly integrate with respect to $x^0$. Multiplying it by $\mathcal{G}$, we obtain the function $\mathcal{G}^4$, which is not a well-defined distribution in the limit of a thin shell. To be able to deal with this term using the method above described, we proceed as the following. Rather than treating $g_{\rm int}$ as a finite valued constant, we treat it as a function of $\sigma_0$ with the condition that $\lim_{\sigma_0\rightarrow 0}g_{\rm int}=0$. Therefore, let us assume that
\begin{equation}
\label{g2d}
g_{\rm int}= \sqrt{2 \pi \ }g_{\rm 2D} \ \sigma_0
\end{equation}
where $g_{\rm 2D}$ is a constant that we refer to as {\it two-dimensional interaction strength}. We can conclude that
\begin{eqnarray}
g_{\rm int}\mathcal{G}^4= g_{\rm 2D} \frac{e^{\frac{(x^0)^2}{\left(\sigma_0/\sqrt{2}\right)^2}}}{\sqrt{\pi}\left(\sigma_0/\sqrt{2}\right)},
\end{eqnarray}
which corresponds to the constant $g_{\rm 2D}$ times a Gaussian distribution with standard deviation $\sigma_0/\sqrt{2}$. With this, the limit below becomes well-defined:
\begin{equation}
\label{rep3delta}
\lim_{\sigma_0\rightarrow 0} g_{\rm int}\mathcal{G}^4
=
g_{\rm 2D} \delta (x^0).
\end{equation}

Now, let us follow the dimensional reduction method integrating equation~(\ref{GP3}) times $\sqrt{(1+\kappa_1 x^0)(1+\kappa_2 x^0)} \ \mathcal{G}(x^0)$ with respect to the variable $x^0$ in the limit of a thin shell:
\begin{eqnarray}
\hspace{-1cm}
\lim_{\sigma_0\rightarrow 0}\int_{-R/2}^{R/2}dx^0\sqrt{(1+\kappa_1x^0)(1+\kappa_2x^0)} \
\mathcal{G}({x^0})
\cdot
\textrm{GP}
\left(
\Psi(x^0,x^1,x^2)
\right)
= 0.
\end{eqnarray}
Evaluating the term~$\textrm{GP}\left(\Psi(x^0,x^1,x^2)\right)$ according to equation~(\ref{GP3}), the above expression becomes
\begin{eqnarray}
\nonumber
\hspace{-2.5cm}
\lim_{\sigma_0\rightarrow 0}\int_{-R/2}^{R/2}dx^0
\cdot
\Bigg\{
-\mu_2\mathcal{G}^2\phi
-\frac{\hbar^2}{8m}
\left(
  \frac{\kappa_1}{1+\kappa_1x^0}
  -
  \frac{\kappa_2}{1+\kappa_2x^0}
\right)^2\mathcal{G}^2\phi
-\frac{\hbar^2}{2m}(\Delta_{\mathcal{M}(x^0)}\phi)\mathcal{G}^2
\\
\ \nonumber
\\
\nonumber
\hspace{-0.5cm}
-\frac{\hbar^2}{2m}\sqrt{(1+\kappa_1x^0)(1+\kappa_2x^0)}
\left
  (\Delta_{\mathcal{M}(x^0)}[(1+\kappa_1x^0)(1+\kappa_2x^0)]^{-1/2}
\right)
\mathcal{G}^2\phi
\\
\ \nonumber
\\
\hspace{-0.5cm}
-\frac{\hbar^2}{m}\sqrt{(1+\kappa_1x^0)(1+\kappa_2x^0)}
g^{ij}
\left(
\partial_i [(1+\kappa_1x^0)(1+\kappa_2x^0)]^{-1/2}\right)
\mathcal{G}^2\left(\partial_j\phi\right)
\\
\ \nonumber
\\
\nonumber
\hspace{-0.5cm}
+\frac{\hbar\omega_2}{\sigma_0^2}(x^0)^2\mathcal{G}^2\phi
+\frac{m\omega_2^2}{2}(x^0)^2\mathcal{G}^2\phi
+\frac{g_{\rm int}\mathcal{G}^4|\phi|^2\phi}{(1+\kappa_1x^0)(1+\kappa_2x^0)}
\Bigg\}=0,
\end{eqnarray}
where we hide the dependence of $\mathcal{G}$ on $x^0$ and of $\phi$, $\kappa_1$ and $\kappa_2$ on $x^1,x^2$ for simplicity.

Interchanging the limit $\lim_{\sigma_0\rightarrow 0}$ with the integral as it is done in equation~(\ref{rep1delta}), and considering equations~(\ref{rep2delta}) and (\ref{rep3delta}), the above expression becomes
\begin{eqnarray}
\nonumber
\hspace{-2.5cm}
\int_{-R/2}^{R/2}dx^0
\Bigg\{
-\mu_2\phi
-\frac{\hbar^2}{8m}
\left(
  \frac{\kappa_1}{1+\kappa_1x^0}
  -
  \frac{\kappa_2}{1+\kappa_2x^0}
\right)^2\phi
-\frac{\hbar^2}{2m}(\Delta_{\mathcal{M}(x^0)}\phi)
\\
\ \nonumber
\\
\nonumber
\hspace{-0.5cm}
-\frac{\hbar^2}{2m}\sqrt{(1+\kappa_1x^0)(1+\kappa_2x^0)}
\left
  (\Delta_{\mathcal{M}(x^0)}[(1+\kappa_1x^0)(1+\kappa_2x^0)]^{-1/4}
\right)
\phi
\\
\nonumber
\\
\hspace{-0.5cm}
\label{SecondRooww}
-\frac{\hbar^2}{m}\sqrt{(1+\kappa_1x^0)(1+\kappa_2x^0)}
g^{ij}
\left(
\partial_i [(1+\kappa_1x^0)(1+\kappa_2x^0)]^{-1/2}\right)
\left(\partial_j\phi\right)
\\
\ \nonumber
\\
\nonumber
\hspace{-0.5cm}
+\frac{\hbar\omega_2}{2}\phi
+\frac{m\omega_2^2}{2}(x^0)^2\phi
+\frac{g_{\rm 2D}|\phi|^2\phi}{(1+\kappa_1x^0)(1+\kappa_2x^0)}
\Bigg\}\delta(x^0)=0,
\end{eqnarray}

To perform this integral, we can attribute $x^0=0$ to all the terms inside the curly brackets. The only terms where this procedure requires more caution are those from second and third rows, because we need to perform the derivatives before replacing the value of $x^0$. One can check that the contribution of this term vanishes, being enough to verify that the derivatives $\partial_i$ and $\partial_i\partial_j$ of the term $[(1+\kappa_1x^0)(1+\kappa_2x^0)]^{-1/4}$ vanish for $x^0=0$. 



With this, the above integral becomes
\begin{eqnarray}
\label{2dpg1}
\hspace{-1cm}
-\mu_2\phi
-\frac{\hbar^2}{8m}
\left(\kappa_1-\kappa_2
\right)^2\phi
-\frac{\hbar^2}{2m}\Delta_{\mathcal{M}}\phi
+\frac{\hbar\omega_2}{2}\phi
+g_{\rm 2D}|\phi|^2\phi=0.
\end{eqnarray}
Denoting
\begin{equation}
\label{Vgeom}
V_{\rm geom}(x^1,x^2)=-\frac{\hbar^2}{8m}[\kappa_1(x^1,x^2)-\kappa_2(x^1,x^2)]^2
\end{equation}
as the geometric potential and
\begin{equation}
V_{\rm ext}(x^1,x^2)=\frac{\hbar\omega_2(x^1,x^2)}{2}
\end{equation}
as the external potential,
we can rewrite equation~(\ref{2dpg1}) as
\begin{equation}
\left(
  -\mu_2
  -\frac{\hbar^2}{2m}\Delta_{\mathcal{M}}
  +V_{\rm geom}
  +V_{\rm ext}
  +g_{\rm 2D}|\phi|^2
\right)\phi=0
\label{maingp}
\end{equation}
where $\phi=\phi(x^1,x^2)$.

This is a 2DGP equation for the two-dimensional wave-function $\phi(x^1,x^2)$, where we obtain two effective potentials $V_{\rm geom}(x^1,x^2)$ and $V_{\rm ext}(x^1,x^2)$. The first potential depends on the curvatures $\kappa_1(x^1,x^2)$ and $\kappa_2(x^1,x^2)$ of the surface $\mathcal{M}$, while the latter can be any function, and it depends on the asymmetries of the second term of the confinement frequency $\omega_2(x^1,x^2)$.

The geometric potential $V_{\rm geom}$ was first developed in the seminal references~\cite{Jensen, daCosta}. In a previous reference~\cite{BECmanifold} of some of us, we also computed this geometric potential, expressed in terms of the determinant of the metric $g_0$. Moreover, we did not explore further consequences of this term in that article, since we focused on applying our methods to a sphere, where the geometric potential vanishes. This happens because the principal curvatures $\kappa_1$ and $\kappa_2$ are equivalent on a sphere.

In the following sections, we are going to apply the 2DGP equation to an ellipsoid and compare the effects generated by the geometric and external potentials on the ground state.

\section{Thomas Fermi approximation} \label{SecTF}

In the three-dimensional GP equation, the Thomas Fermi approximation must be taken when the interactions $g_\mathrm{int}$ are strong. However, we assume that $g_\mathrm{int}$ are weak. In fact, according to equation~(\ref{g2d}), we have that $g_\mathrm{int}\rightarrow 0$ in the limit of a thin shell. Thus, within the model we are working on, we cannot use Thomas-Fermi methods to solve equation~(\ref{GP}).

On another side, in the 2DGP equation~(\ref{maingp}), the two-dimensional interactions are assumed to be the constant value $g_\mathrm{2D}$. Considering this constant to be strong, we can apply the Thomas Fermi solution to the 2DGP equation, and the ground state density is well-approximated by
\begin{eqnarray}\label{TFsolution}
\hspace{-1cm}
|\phi_0(x^1,x^2)|^2=\frac{1}{g_{\rm 2D}}
\left(
\mu_2
-\frac{\hbar\omega_2(x^1,x^2)}{2}
+\frac{\hbar^2}{8m}[\kappa_1(x^1,x^2)-\kappa_2(x^1,x^2)]^2
\right)
\end{eqnarray}
If $\omega_2$ is a constant, then the ground state is fully defined by the difference between the mean curvatures. The higher density happens in the regions of the highest difference of curvatures $\kappa_ 1$ and $\kappa_2$.

With this, one can conclude that, in the absence of external asymmetries, geometric properties of the surface $\mathcal{M}$ can be directly observed by measuring the ground state of a Bose-gas confined in this area.

In the following sections, we will analyze the situation where the surface $\mathcal{M}$ is an ellipsoid, which is motivated by the bubble trap experiments.

\section{Ellipsoid} \label{SecEllipsoid}

Motivated by the current experiments to implement a bubble trap, we will apply the above formalism for a BEC confined on the surface of a prolate ellipsoid given by the following equation
\begin{equation}
\frac{x^2}{a^2}+\frac{y^2}{a^2}+\frac{z^2}{b^2}=R^2
\label{ellipsoid}
\end{equation}
in Cartesian coordinates, where $a$ and $b$ are dimensionless.

To study the ground state, we need to express the ellipsoid in two variables $x^1$ and $x^2$. To do that, we first consider the prolate spheroidal coordinates $(r,\varphi,\theta)$ to describe the three-dimensional space. It is given by the transformation
\begin{eqnarray}
&x=ra\cos\theta \sin\varphi \nonumber\\
&y=ra\sin\theta \sin\varphi \label{prolatecoord}\\
&z=rb\cos\varphi \nonumber
\end{eqnarray}
where $r>0$, $\varphi\in[0,\pi]$, and $\theta\in[0,2\pi)$.
A constant value of $r=R$ defines the ellipsoid of equation~(\ref{ellipsoid}). Thus, we can describe the ellipsoid with the two variables $x^1=\varphi$ and $x^2=\theta$.

\begin{figure}
\includegraphics[scale=0.109]{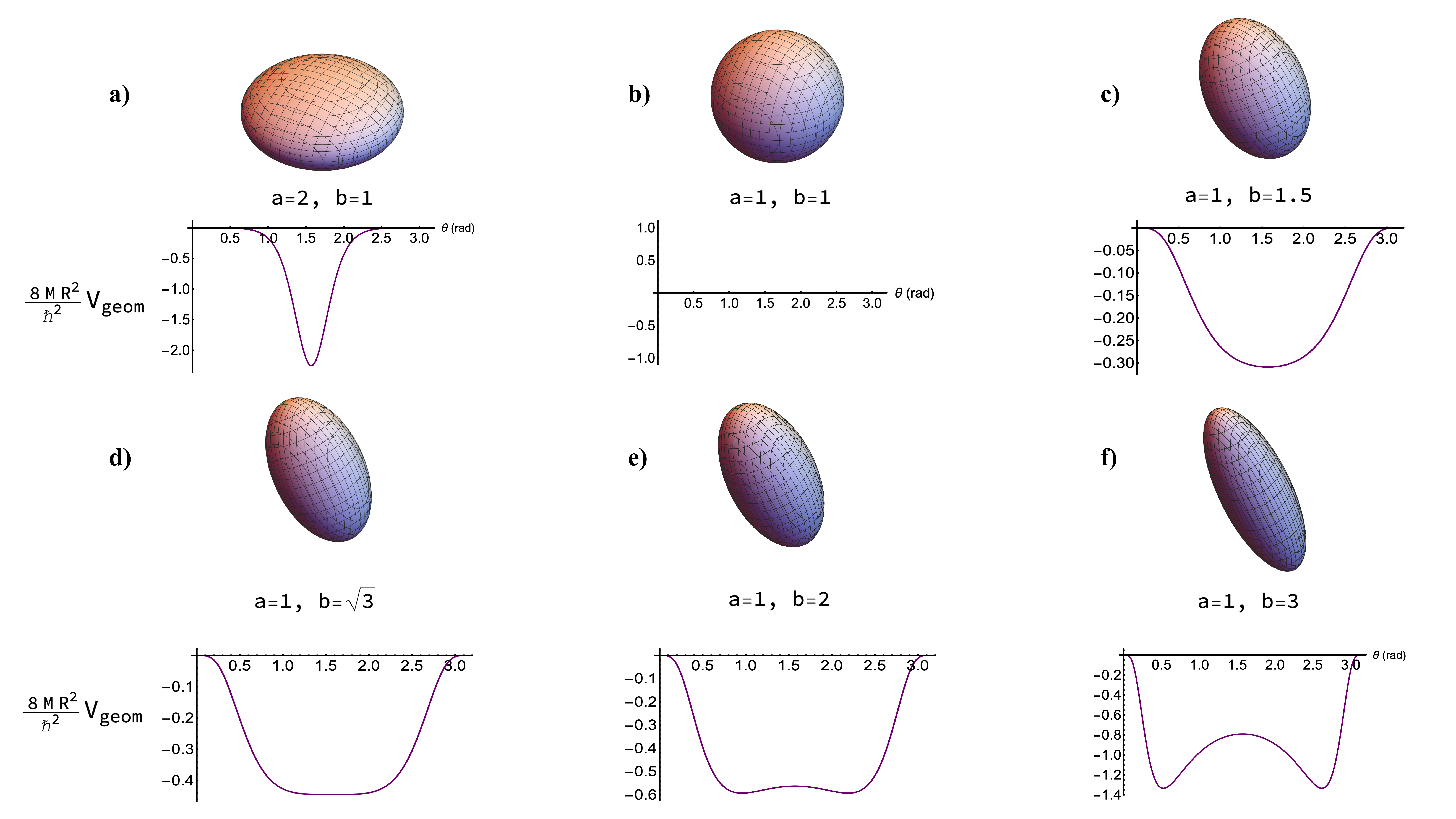}
\caption{\label{fig1}
Geometric potential on ellipsoids for different values of $a$ and $b$.}
\end{figure}

The mean and Gaussian curvatures for such an ellipsoid are given by
\begin{equation}
H=\frac{b}{2a}
\frac{2a^2+(b^2-a^2)\sin^2\varphi}{(a^2\cos^2\varphi+b^2\sin^2\varphi)^{3/2}}
\end{equation}
and
\begin{equation}
K=\frac{b^2}{(a^2\cos^2\varphi+b^2\sin^2\varphi)^{2}}.
\end{equation}
Using the relation~(\ref{kappaHK}), we obtain that $(\kappa_1-\kappa_2)^2=H^2-K$, and from the above formulas, the geometric potential is given by
\begin{equation}
V_\mathrm{geom}=-\frac{\hbar^2}{8mR^2}\frac{b^2(a^2-b^2)^2\sin^4\varphi}{a^2[a^2\cos^2\varphi+b^2\sin^2\varphi]^3}.
\end{equation}
Note that there is no dependence on $\theta$, which is expected due to the symmetry of the ellipsoid. In figure~\ref{fig1}, we plot the dependence of the geometric potential with respect to $\varphi$, for different values of $a$ and $b$.

For $a>b$, we have an oblate ellipsoid and the minimum of the potential is always on $\varphi=\pi/2$, that is, on the equator of the ellipsoid. For $a=b$, we have the special case of a sphere, where the geometric potential vanishes, since $\kappa_1=\kappa_2$ everywhere. For $b>a$ we have a prolate ellipsoid, and a minimum appears on the equator for $b/a\leq\sqrt{3}$, and a local maximum for $b/a>\sqrt{3}$. Notably, for any value of the ratio between $a$ and $b$, a global maximum appears on the poles.

Finally, the ground state of the BEC trapped in such an ellipsoid with $\omega_2=0$ is given by 
\begin{eqnarray}
|\phi_0|^2=\frac{1}{g_{\rm 2D}}
\left(
\mu_2
+\frac{\hbar^2}{8mR^2}\frac{b^2(a^2-b^2)^2\sin^4\varphi}{a^2(a^2\cos^2\varphi+b^2\sin^2\varphi)^3}
\right).
\label{groundEllipsoid}
\end{eqnarray}
Note that the denominator $a^2\cos^2\varphi+b^2\sin^2\varphi$ is always positive.
Since $\sin\varphi$ vanishes for $\varphi=0,\pi$, we can directly see that the points with the lowest density are the poles of the ellipsoid.

Comparing this result with those from the literature we find a discrepancy. The numerical simulations for the bubble trap~\cite{LucaMonteCarlo} and even the experiment~\cite{BubblesSpace} show a higher accumulation of atoms on the poles of the ellipsoid, while our results show this higher accumulation on the equator. We analyse the reasons for this difference in the following section.

\section{Ground state and comparison to the bubble trap} \label{SecBubble}

There is a discrepancy between the ground state evaluated for the bubble trap and the one we found in the previous section. One of the reasons for this is that the potential in the experiments does not satisfy one of our hypotheses: the dominant term $\omega_0$ of the confinement frequency is not constant. To see this, consider the bubble-trap potential, which was derived in references~\cite{OZobayPRL,OZobayPRA}
\begin{equation}\label{GarrawayPot1}
U(x,y,z)=\hbar M_F\Delta\sqrt{
	\left(
		\frac{m}{4\hbar\Delta}
(\omega_x^2x^2+\omega_y^2y^2+\omega_z^2z^2)
		-1
	\right)^2
+\frac{\Omega^2}{\Delta^2} \ },
\end{equation}
which is written is Cartesian coordinates $(x,y,z)$. The parameters $\Delta$, $\Omega$, and $m$ are the detuning of the radio frequency magnetic field, the Rabi frequency between the hyperfine levels and the atomic mass respectively, while $M_F$ labels the highest dressed state. Moreover, it is assumed that $\Delta\gg\Omega$.

The minimum of this potential defines an ellipsoid with the equation:
\begin{equation}
  \omega_x^2x^2+\omega_y^2y^2+\omega_z^2z^2
=\frac{4\hbar\Delta}{m}
\end{equation}
If $\omega_x=\omega_y$, we obtain a symmetric ellipsoid as in equation (\ref{ellipsoid}). In this case, we can set
\begin{equation}\label{ellipsoidab}
a=1
\ , \ \ \ \
b=\frac{\omega_x}{\omega_z}
\ , \ \ {\rm and} \ \ \ \
R=\sqrt{\frac{4\hbar\Delta}{m\omega_x^2}}
\end{equation}
and the potential~(\ref{GarrawayPot1}) can be re-expressed as
\begin{equation}
\label{UbubbleCart}
U(x,y,z)=\hbar M_F\Delta\sqrt{\frac{m\omega_x}{4\hbar\Delta}
	\left(
      \frac{x^2}{a^2}
      +\frac{y^2}{a^2}
      +\frac{z^2}{b^2}
      -R^2
	\right)^2
+\frac{\Omega^2}{\Delta^2} \ }
\end{equation}
In prolate spheroidal coordinates, it can be written as
\begin{equation}
\label{UbubbleProlSphe}
U(r)=\hbar M_F\Delta\sqrt{\frac{m\omega_x}{4\hbar\Delta}(r^2-R^2)^2
+\frac{\Omega^2}{\Delta^2} \ }
\end{equation}
which can be Taylor expanded as
\begin{equation}\label{U72}
U(r)\simeq \hbar M_F\Omega+
\frac{M_F\Delta m\omega_x}{\Omega}(r-R)^2+...
\end{equation}
In Gaussian normal coordinates the potential~(\ref{GarrawayPot1}) becomes approximately
\begin{equation}\label{U73}
U(x^0,\varphi)\simeq M_F\hbar\Omega+
\frac{m\omega_0^2(\varphi)}{2}(x^0)^2+...
\end{equation}
where
\begin{eqnarray}\label{omegabbtrap}
\omega_0^2(\varphi)=
\frac{2M_F\omega_x}{a^2b^2}\frac{\Delta}{\Omega}(a^2\cos^2\varphi+b^2\sin^2\varphi).
\end{eqnarray}
This expansion is developed in~\ref{AppBubbleTrap}. 

Remember that $\Delta\gg\Omega$, thus, we cannot apply the dimensional reduction developed in the previous sections to this kind of potential, since one of our hypotheses is that $\omega_0$ is constant.
Note that this is not a specific restriction of our dimensional reduction method. In fact, it is not possible to define a two-dimensional wave function in the limit of a thin surface for this potential~(\ref{GarrawayPot1}), except for spheres and small deformations on a sphere~\cite{QuasiEllipsoid}, where there is only an infinitesimal difference between the parameters $a$ and $b$.

Here, we work with an ellipsoid chosen in advance, with a finite difference between the parameters $a$ and $b$. For this kind of surface, it is not possible to perform a well-defined dimensional reduction considering the bubble trap potential~(\ref{GarrawayPot1}). Even though, we still can use our methods to find some comparison. Motivated by the formula of equation (\ref{omegabbtrap}), we will consider instead the following confinement
\begin{equation}
U(x^0,\varphi)=\frac{m\left[\omega_0+\omega_2(\varphi)\right]^2}{2}(x^0)^2
\end{equation}
with $\omega_0$ defined as in section~\ref{SecModel}, and
\begin{equation}
\omega_2(\varphi)=\beta\frac{\hbar}{4mR^2}(
a^2\cos^2\varphi+b^2\sin^2\varphi),
\end{equation}
where $\beta$ is a dimensionless constant, with $\mathcal{O}(\beta)=\mathcal{O}(1)$. Thus, we also have that $\mathcal{O}[\omega_2(\varphi)/\omega_R]=\mathcal{O}(1)$ for all $\varphi\in[0,\pi]$, satisfying the conditions of the method developed in the previous sections.

We can then apply the 2DGP equation for this kind of confinement. The ground state from the Thomas Fermi solution becomes then:
\begin{eqnarray}
\hspace{-2.5cm}|\phi_0|^2=\frac{1}{g_{\rm 2D}}
\left(
\mu_2
-\beta\frac{\hbar^2}{8mR^2}(a^2\cos^2\varphi+b^2\sin^2\varphi)
+\frac{\hbar^2}{8mR^2}\frac{b^2(a^2-b^2)^2\sin^4\varphi}{a^2(a^2\cos^2\varphi+b^2\sin^2\varphi)^3}
\right)
\end{eqnarray}

We plot the atom density in figure~\ref{fig2} for $\beta=0$, $\beta=0.2$, $\beta=0.5$, and $\beta=1$, considering $a=1$ and $b=2$. For $\beta=0$ we find a higher accumulation of atoms on the equator, that is, at $\varphi=\pi/2$, while for $\beta=0.5$ and $\beta=1$ we observe a higher accumulation of atoms on the poles, that is, at $\varphi=0$ and $\varphi=\pi$. For $\beta=0.2$, we have an intermediate situation, where the equator and poles both show a higher accumulation in comparison with complementar regions.

 \begin{minipage}{\linewidth}
      \centering
      \begin{minipage}{0.46\linewidth}
          \begin{figure}[H]
           \hspace{-1cm}   \includegraphics[scale=0.32]{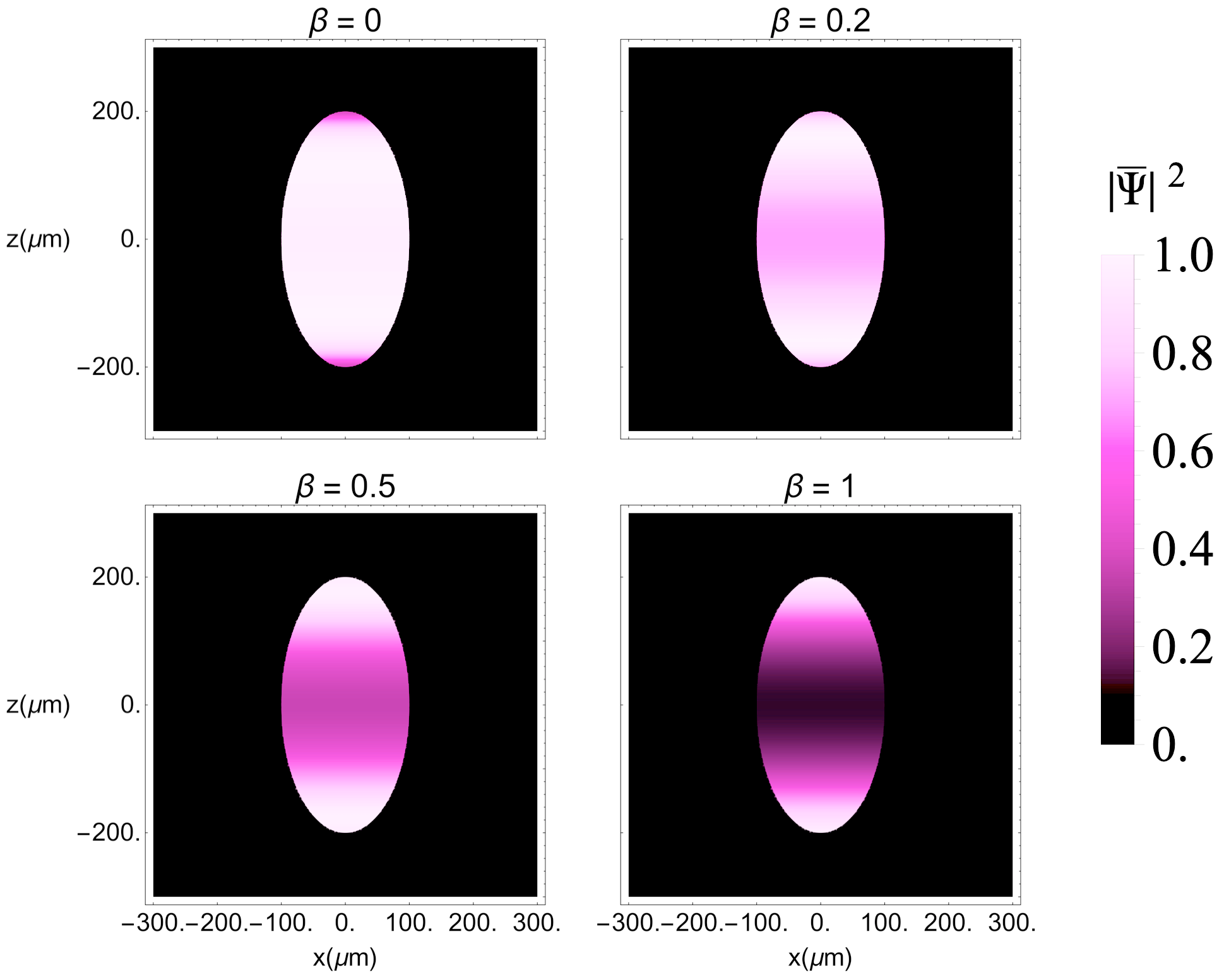}
              \caption{Profile of the gas density on the ellipsoid surface for different values of $\beta$.}
              \label{fig2}
          \end{figure}
      \end{minipage}
      \hspace{0.05\linewidth}
      \begin{minipage}{0.46\linewidth}
          \begin{figure}[H]

              \vspace{20mm}
              
              \hspace{3cm} \includegraphics[scale=1.3]{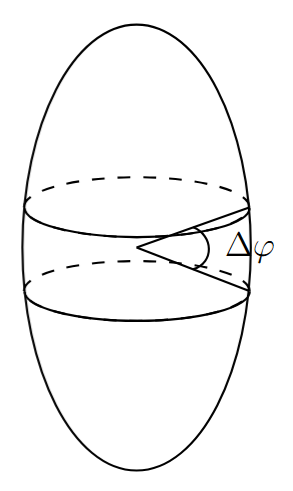}

              \vspace{10mm}
              
              \caption{Strip on an ellipsoid, being definined by $\varphi\in [(\pi-\Delta\varphi)/2,(\pi+\Delta\varphi)/2]$ and $\theta\in[0,2\pi)$.}
              \label{Strip}
          \end{figure}
      \end{minipage}
  \end{minipage}

\

\

To gain a better notion on these differences, we compute the number of atoms on a strip along the equator of angle $\Delta\varphi$, as illustrated in figure~\ref{Strip}. More precisely, we compute
\begin{equation}
N_{\Delta\varphi}=\int_0^{2\pi}\int_{\frac{\pi-\Delta\varphi}{2}}^{\frac{\pi-\Delta\varphi}{2}} |\phi_0(\varphi)|^2 dA.
\end{equation}
where $\theta$ ranges from $0$ to $2\pi$, $\varphi$ ranges from $(\pi-\Delta\varphi)/2$ to $(\pi+\Delta\varphi)/2$, and $dA=R^2a\sin\varphi\sqrt{a^2\cos^2\varphi+b^2\sin^2\varphi \ }d\theta d\varphi$ is the area element.

\begin{figure}
{\bf a)} \hspace{7.8cm} {\bf b)} \hspace{7.4cm}

\includegraphics[scale=.2]{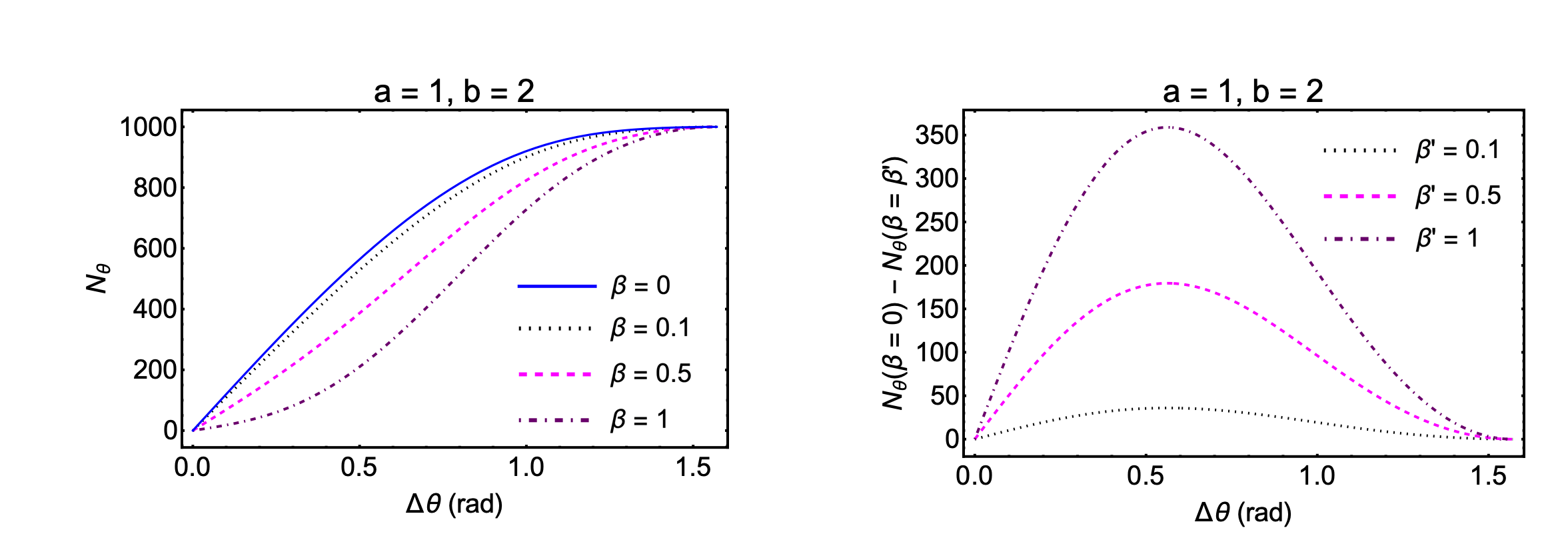}
\caption{\label{fig7}
{\bf a)} Number of atoms for different values of $\beta$ in the interval $\Delta\varphi$.
{\bf b)} Difference on the atom number between $\beta=0$ and $\beta'$ in the interval $\Delta\varphi$, for different values of $\beta'$.
}
\end{figure}

We plot this atom number in figure~\ref{fig7}{\bf a} for the different values of $\beta$ that we have already considered, where $N=1000$ is the total atom number. The differences of the atom number in a strip between the uniform case, where $\beta=0$, and $\beta'$, for $\beta'=0.1$, $\beta'=0.5$ and $\beta'=1$, is plotted in figure~\ref{fig7}{\bf b}. We see that the biggest differences happen for $\Delta\varphi$ around $\pi/6$ in all three cases, and that they can reach an amount of hundreds of atoms. This shows that the difference in the atom number for different confinements is eligible to be experimentally measured.

\section{Discussion} \label{SecDisc}

As we have shown in section~\ref{SecBubble}, it is not possible to implement a uniform two-dimensional confinement with the bubble-trap potential~(\ref{GarrawayPot1}), since its asymmetry diverges in the thin limit confinement. On the one hand, without a significant modification to the current techniques to realize a bubble trap, it would be quite challenging to observe the geometric properties on the ground state of a gas confined to the surface of an ellipsoid. On the other hand, these asymmetries could be quite useful for different purposes. For instance, in~\cite{ExpGuo}, the asymmetries are conveniently employed in the realization of a Bose gas on an oblate ellipsoid in a terrestrial laboratory. These asymmetries lead to an effective potential on a quasi 2D bubble and are strong enough to partially compensate the gravitational sag effect.

However, a conceptual understanding of geometric effects in a bubble trap is still lacking. For instance, reference~\cite{Perspectives2} mentions possible effects of curvature on an ellipsoid, stating that the curvature can lead to local confinement of atoms near the higher curvature regions. Even though this statement is not strictly incorrect, it deserves caution. 

According to the analysis we performed here, for an oblate ellipsoid, geometric effects indeed lead to a higher accumulation in the region of higher Gaussian curvature, which is the equator. This can be checked considering equation~(\ref{groundEllipsoid}), with $a>b$, or simply noticing figure~\ref{fig1}a. This coincidence happens in this particular case because the region of higher curvature is the same as the region of higher difference between the principal curvatures.

However, for a prolate ellipsoid, which is the case we focused on in this paper, we have the opposite behavior. The regions of higher Gaussian curvature are the poles, while the region of higher atomic accumulation due to geometric effects is the equator. This contrasts with the inference mentioned above from~\cite{Perspectives2}.

We recall that the word {\it curvature} is generally used to denote the Gaussian curvature, introduced in equation~(\ref{GaussianCurv}), which can also be referred to as {\it intrinsic curvature}. This curvature receives considerably more attention than the mean and principal curvatures, introduced in equations~(\ref{meanCurv}) and~(\ref{kappaHK}), which can also be referred to as {\it extrinsic curvatures}. The reason for this is that, as a consequence of the {\it Gauss' Egregium Theorem}~\cite{doCarmo}, the Gaussian curvature is an intrinsic two-dimensional geometric property of a surface.

For instance, one can consider only the inner angles of a local triangle to compute the Gaussian curvature at a point. In particular, if their sum is equal to $\pi$, then the Gaussian curvature vanishes and the surface is locally flat at that point. If their sum is larger than $\pi$, then the curvature is positive, and if it is smaller than $\pi$, the curvature is negative. 
In contrast, calculating the principal and mean curvatures requires taking into account the three-dimensional space in which the surface is embedded. For instance, one must examine how the normal vector to a surface varies along a path on this same surface. By analyzing the variation of normal vectors, we are inherently accessing tree-dimensional properties of the Euclidean space.

Observe that the geometric potential in equation~(\ref{Vgeom}) depends on the principal curvatures, and consequently, the ground state~(\ref{TFsolution}) also depends on them. The effects found here are understood as geometric effects of a surface, which is a two-dimensional manifold embedded in a three-dimensional space. Although it is correct to interpret the gas as occupying a two-dimensional region in the limit of a thin shell, where $\sigma_0\rightarrow\infty$, it still exhibits properties of a system that resides in a three-dimensional space.

\section{Conclusions} \label{SecConc}

In this paper, we unravel the geometric meaning of the ground state of a BEC confined on a surface. To do that, we compute the ground state using a perturbative expansion on the first orders of magnitude of the width of the surface, followed by dimensional reduction techniques. While one can find a term in the expression of the ground state density that appears exclusively due to the curvatures of the surface, there is also another term that does not depend directly on the geometry, but on the asymmetries of confinement. These asymmetries depend on the engineering used to create the confinement potential and might indirectly depend on the geometry due to technical reasons.

In particular, we show that the accumulation of atoms on the poles of a bubble trap happens to be exactly the latter case. It depends on the experimental techniques used and not on the geometrical properties of an ellipsoid. Instead, we show that if these technical asymmetries were eliminated or at least made weaker, it would be possible to observe an accumulation of atoms on the equator of an ellipsoid. This latter accumulation happens exclusively due to the geometric properties of this surface. 

The methods used in this paper for deriving the 2DGP equation are elementary, being necessary only differential and integral calculus and basic knowledge of Dirac delta function distribution. While reference~\cite{BECmanifold} considers the minimization principles for Action and Energy, here we only need to assume that the GP equation holds.

For solving the 2DGP equation, we used the Thomas-Fermi approximation, which suits well for our model, since the two-dimensional interactions $g_{2D}=g_{\rm int}/(2\sqrt2 \ \sigma_0)$  are strong for small $\sigma_0$. For future research, it is worth exploring the role of the Laplace-Beltrami operator $\Delta_\mathcal{M}$ in the ground state or even in the dynamics of a BEC uniformly confined on an ellipsoid. Further interesting applications of our results can be explored by considering different geometries, such as cylindrical and toroidal shapes~\cite{Tononinew, Nikolaieva_2023, 10.1116/5.0211426}, or even adapting them to deal with non-orientable surfaces, like a M\"obius strip~\cite{Mobius}. 

\section{Acknowledgements}

S.M.O. acknowledges the fund support from DAAD-CAPES, project PROBAL 88887.899314/2023-00, and FAPESP, project 2020/05057-4.
N.S.M. acknowledges the projects INDORBLE 09I03-03-V04-00679 for excellent researchers R2, DeQHOST APVV-22-0570, and QUAS VEGA 2/0164/25. N.S.M. has benefited from the activities of COST Action CA23115:
Relativistic Quantum Information, funded  by COST (European Cooperation
in Science and Technology).

S.M.O. and N.S.M. thank Axel Pelster for the valuable discussions and for his support throughout all stages of this work. S.M.O. thanks F. Ednilson A. dos Santos for all the support during the development of this work. The authors acknowledge Hélène Perrin for useful discussions.

\appendix

\section{Justification for the term $\mu_1=0$} \label{Appmu1}

In the main text, we obtained the equation~(\ref{checkzero2}), which can be re-expressed as
\begin{equation}
\frac{\partial S_1}{\partial x^0}
=\frac{\partial \ln\sqrt{(\det g)^{-1/2}}}{\partial x^0}
+\frac{\mu_1m\sigma_0^2}{\hbar^2x^0}
\end{equation}
Integrating on $x^0$, we obtain
\begin{equation}
S_1
=\ln\sqrt{(\det g)^{-1/2}}
+\frac{\mu_1m\sigma_0^2}{\hbar^2}\ln|x^0|+\bar{S_1}(x^1,x^2)
\end{equation}
where $\bar{S_1}(x^1,x^2)$ is a parameter that is constant with respect to $x^0$. Remember that $\ln|x^0|$ diverges for $x^0\rightarrow0$, which means that the factor $S_1$ would diverge on the surface.
Since this solution has no physical meaning, we must set $\mu_1=0$.

\section{Gaussian distributions and Dirac delta function} \label{AppInteg}

Here, we express well-known integrals of a Gaussian distribution, which is the square of the Gaussian wave-function~(\ref{solutionHarmOsc}), and of this Gaussian distribution times a polynomial~\cite{Gradshteyn}. We use it to derive equation~(\ref{rep2delta}) of the main text.

Indeed, one can show that
\begin{equation}
\label{int1}
\int_{-\infty}^{\infty}\mathcal{G}^2(x^0)dx^0
=1;
\end{equation}



\noindent Moreover, for a natural even number $n$, we have that
\begin{equation}\label{intGxnequal}
\int_{-\infty}^{\infty}(x^0)^n\mathcal{G}^2(x^0)dx^0
=(n-1)!!\frac{\sigma_0^{n}}{2^{n/2}},
\end{equation}

\noindent while for $n$ odd, we have that
\begin{equation}\label{intGxnoddequal}
\int_{-\infty}^{\infty}(x^0)^n\mathcal{G}^2(x^0)dx^0
=0.
\end{equation}
The integration limit in the above equations ranges from $-\infty$ to $\infty$, while in the main text, the integrations were evaluated from  $-R/2$ to $R/2$. Since we are considering $\sigma_0\ll R$, we have approximately that~\cite{Gradshteyn}:
\begin{equation}\label{intGxn}
\int_{-R/2}^{R/2}(x^0)^n\mathcal{G}^2(x^0)dx^0
=(n-1)!!\frac{\sigma_0^{n}}{2^{n/2}}+\mathcal{O}(e^{-R/\sigma_0})
\end{equation}
for $n$ even, and
\begin{equation}\label{intGxnodd}
\int_{-R/2}^{R/2}(x^0)^n\mathcal{G}^2(x^0)dx^0
= 0 +\mathcal{O}(e^{-R/\sigma_0})
\end{equation}
for $n$ odd.

Now we are going to show that equation~(\ref{rep2delta}) is indeed valid. Let $\delta_{\epsilon}(x^0)$ be a family of integrable functions. We say that
\begin{equation}
\label{limitDelta}
\lim_{\epsilon\rightarrow 0}\delta_{\epsilon}(x^0)=\delta(x^0)
\end{equation}
where $\delta(x^0)$ is the Dirac delta, if and only if
\begin{equation}
\label{defDelta}
\lim_{\epsilon\rightarrow 0}\int_{\mathcal{N}\{x^0=0\}} f(x^0) \delta_{\epsilon}(x^0)dx^0=f(0),
\end{equation}
for all well-behaved functions $f(x^0)$, where $\mathcal{N}\{x^0=0\}$ is a neighbourhood around $x^0=0$. As in the main text, for simplicity, we consider that $\mathcal{N}\{x^0=0\}=[-R/2,R/2]$.
Remember that the Dirac delta is not a function, and that the limit~(\ref{limitDelta}) is not rigorously a limit, but a convention of notation.
With this convention, let us show that $\lim_{\sigma_0\rightarrow 0} [(x^0)^2/\sigma_0^2]\mathcal{G}^2=\delta(x^0)/2$, {\it i.e.}, that equation~(\ref{rep2delta}) holds.

Considering the Taylor expansion of a well-behaved function $f(x^0)$ around $x^0=0$, we can write that
\begin{equation}
\int_{-R/2}^{R/2} f(x^0)\frac{(x^0)^2}{\sigma_0^2}\mathcal{G}^2(x^0)
=
\sum_{n=0}^\infty \frac{f^{(n)}(0)}{n! \ \sigma_0^2}\int_{-R/2}^{R/2}(x^0)^{n+2}\mathcal{G}^2(x^0) dx^0.
\end{equation}
From equations~(\ref{intGxn}) and (\ref{intGxnodd}), it becomes
\begin{equation}
=\sum_{n {\rm \ even}} \frac{(n+1)!!}{n! \ 2^{(n+2)/2}}f^{(n)}(0)\sigma_0^n+\mathcal{O}(e^{-R/\sigma_0})= \frac{f(0)}{2}+\frac{3}{8}f''(x^0)\sigma_0^2+...
\end{equation}
From the above equation, we can see that its limit when $\sigma_0\rightarrow 0$ is equal to $f(0)/2$. With this result and the definition~(\ref{limitDelta})--(\ref{defDelta}), we guarantee the validity of equation~(\ref{rep2delta}).

\section{Justification for the vanishing of $S_2$-involved terms after integration
} \label{AppS2}

In section~\ref{PertExp}, we mentioned that the terms associated with $S_2$ vanish within the range we are considering. In this appendix,  we provide the calculations supporting this assertion. However, for a clearer understanding, we recommend reading this appendix after completing section \ref{SecDimRed}.

The terms that would appear in the GP equation~(\ref{GP1}) associated with $S_2$ that are of order $\mathcal{O}(1)$, are
\begin{equation}
-\frac{\hbar}{2m}(2S'_0S'_2+S''_2).
\end{equation}
This is because, according to the hypotheses imposed to the Ansatz~(\ref{ansatz0}), we have that $\mathcal{O}(S'_0S'_2)=\mathcal{O}([\sigma_0/R]^{-1}\cdot[\sigma_0/R])=\mathcal{O}(1)$ and $\mathcal{O}(S''_2)=\mathcal{O}(1)$.

Following the methods performed in sections~\ref{PertExp} and \ref{SecDimRed}, we need to show that 
\begin{eqnarray}\label{IntS0S2}
\lim_{\sigma_0\rightarrow 0}\int_{-R/2}^{R/2} (2S_0S_2'+S_2'') \mathcal{G}^2dx^0=0.
\end{eqnarray}
From equation~(\ref{S0'}), we have that:
\begin{eqnarray}
2S_0S_2'+S_2''=-\frac{2x^0 S_2'}{\sigma_0^2}+S_2''.
\end{eqnarray}
Let us consider that $S_2$ can be Taylor expanded around $x^0=0$. With this, we have that $S_2$ is a sum of terms involving $(x^0)^n$, $n\in\mathds{N}$. If we show that
\begin{eqnarray}
\lim_{\sigma_0\rightarrow 0}\int_{-R/2}^{R/2} \left(-\frac{2x^0}{\sigma_0^2}([x^0]^n)'+([x^0]^n)''\right) \mathcal{G}^2dx^0=0,
\end{eqnarray}
then we guarantee that equation (\ref{IntS0S2}) holds. Evaluating the above expression, we have that
\begin{eqnarray}
\lim_{\sigma_0\rightarrow 0}\int_{-R/2}^{R/2} \left(-\frac{2n(x^0)^n}{\sigma_0^2}+n(n-1)(x^0)^{n-2}\right) \mathcal{G}^2dx^0.
\end{eqnarray}
From equation~(\ref{intGxn}), we evaluate that the above becomes approximately
\begin{eqnarray}
\lim_{\sigma_0\rightarrow 0} 
\left(
-\frac{2n(n-1)!!\sigma_0^n}{2^{n/2}\sigma_0^2}+\frac{n(n-1)(n-3)!!\sigma_0^{n-2}}{2^{(n-2)/2}}
\right),
\end{eqnarray}
and we can directly check that the terms inside the parentheses vanish, as we would like to show.

With this, we conclude that the terms that appear in the three-dimensional GP equation that are associated with $S_2$, vanish in the 2DGP equation up to order $\mathcal{O}(1)$.

\section{The Gaussian normal coordinate system for a prolate ellipsoid}\label{AppTaylorExp}

In this appendix, we show how to transform between the Cartesian, prolate spheroidal, and Gaussian normal coordinate systems, with the surface~$\mathcal{M}$ being the prolate ellipsoid defined in equation~(\ref{ellipsoid}).

To work on these formulas, we initially express a generic point ${\bf q}$ and the point ${\bf p}$ on the ellipsoid, which is the closest one to ${\bf q}$ on this surface. Due to the symmetry of the prolate ellipsoid, we have that the angle $\theta$ in prolate spheroidal coordinates~(\ref{prolatecoord}) for both ${\bf p}$ and ${\bf q}$ is the same. Without loss of generality, we can consider that $\theta=0$ (equivalently, $y=0$) and work only on the plane defined by this constraint (equivalently, plane $x-z$). Figure \ref{EllipsoidThetaCte} illustrates this ellipse, which is the intersection of the original ellipsoid and the plane $x-z$, together to and the points ${\bf p}$ and ${\bf q}$.

For the point ${\bf p}$, in the modified spherical coordinates we have $r=R$ and in the Gaussian normal coordinates we have $x^0=0$. With that said, denote the Cartesian coordinates of ${\bf p}$ and ${\bf q}$ as $(x_p,z_p)$ and $(x,z)$, the modified spherical coordinates as $(R,\varphi)$ and $(r,\overline{\varphi})$, and the Gaussian normal coordinate system as $(0,\varphi)$ and $(x^0,\varphi)$, respectively. Note that in the Gaussian normal coordinate system we are choosing $x^2=\theta$. See figure~\ref{EllipsoidThetaCte} for an illustration of these coordinates. For simplicity, we are hiding the third coordinate in all the cases, that is, we will omit $y$, $\theta$ and $x^2$. Due the above discussion, we were free to set conveniently the values of $y$ and $\theta$, and the same argument holds to $x^2$.

From the definition of Gaussian normal coordinate system, we have the relation
\begin{equation}\label{pandq}
	{\bf q}={\bf p}+x^0{\bf n},
\end{equation}
where ${\bf n}$ is the normal vector to the ellipsoid at the point ${\bf p}$. This expression can be visualized in figure~\ref{EllipsoidThetaCte}. 

Note that we can express the variables from different coordinate systems as
\begin{eqnarray}\label{coordinates}
x_p=Ra\sin\varphi \hspace{2cm}
&x=ra\sin\overline{\varphi}
\\
z_p=Rb\cos\varphi \hspace{2cm}
&z=rb\cos\overline{\varphi}
\nonumber
\end{eqnarray}

\tikzset{every picture/.style={line width=0.75pt}} 
\begin{figure}
\begin{center}
\begin{tikzpicture}[x=0.75pt,y=0.75pt,yscale=-1,xscale=1]

\draw   (45.76,265.93) .. controls (45.76,158.51) and (105.76,71.43) .. (179.76,71.43) .. controls (253.77,71.43) and (313.76,158.51) .. (313.76,265.93) .. controls (313.76,373.35) and (253.77,460.43) .. (179.76,460.43) .. controls (105.76,460.43) and (45.76,373.35) .. (45.76,265.93) -- cycle ;
\draw  [dash pattern={on 4.5pt off 4.5pt}]  (179.76,71.43) -- (179.76,265.93) ;
\draw  [color={rgb, 255:red, 0; green, 0; blue, 0 }  ,draw opacity=1 ][fill={rgb, 255:red, 0; green, 0; blue, 0 }  ,fill opacity=1 ] (257,110.21) .. controls (257,108.68) and (258.25,107.43) .. (259.79,107.43) .. controls (261.32,107.43) and (262.57,108.68) .. (262.57,110.21) .. controls (262.57,111.75) and (261.32,113) .. (259.79,113) .. controls (258.25,113) and (257,111.75) .. (257,110.21) -- cycle ;
\draw    (259.79,110.21) -- (179.76,265.93) ;
\draw    (350.76,49.19) -- (179.76,265.93) ;
\draw  [color={rgb, 255:red, 0; green, 0; blue, 0 }  ,draw opacity=1 ][fill={rgb, 255:red, 0; green, 0; blue, 0 }  ,fill opacity=1 ] (348.76,49.19) .. controls (348.76,47.65) and (350.01,46.4) .. (351.55,46.4) .. controls (353.08,46.4) and (354.33,47.65) .. (354.33,49.19) .. controls (354.33,50.73) and (353.08,51.97) .. (351.55,51.97) .. controls (350.01,51.97) and (348.76,50.73) .. (348.76,49.19) -- cycle ;
\draw    (350.76,49.19) -- (259.79,110.21) ;
\draw    (179.76,168.68) .. controls (198.76,157.43) and (215.76,163.43) .. (222.76,182.43) ;
\draw    (179.76,206.43) .. controls (192.76,195.19) and (212.76,196.19) .. (220.76,214.19) ;

\draw (197,148) node [anchor=north west][inner sep=0.75pt]    {$\varphi $};
\draw (199,182) node [anchor=north west][inner sep=0.75pt]    {$\overline{\varphi }$};
\draw (253,85) node [anchor=north west][inner sep=0.75pt]   [align=left] {\textbf{p}};
\draw (354.79,33.43) node [anchor=north west][inner sep=0.75pt]   [align=left] {\textbf{q}};
\draw (287,68) node [anchor=north west][inner sep=0.75pt]    {$x^{0}$};
\draw (183,114) node [anchor=north west][inner sep=0.75pt]    {$\sqrt{x^{2}_p +z^{2}_p}$};
\draw (234,185) node [anchor=north west][inner sep=0.75pt]    {$\sqrt{x^{2} +z^{2}}$};
\draw (388,206) node [anchor=north west][inner sep=0.75pt]   [align=left] {Cartesian:};
\draw (388,270) node [anchor=north west][inner sep=0.75pt]   [align=left] {Prolate spheroidal:};
\draw (388,334) node [anchor=north west][inner sep=0.75pt]   [align=left] {Gaussian normal:};
\draw (460,238) node [anchor=north west][inner sep=0.75pt]   [align=left] {\textbf{p}$\displaystyle =(x_p,z_p)$ \ \ \ \ \ \textbf{q }$\displaystyle =(x ,z)$};
\draw (460,366) node [anchor=north west][inner sep=0.75pt]   [align=left] {\textbf{p}$\displaystyle =(0,\varphi )$ \ \ \ \ \ \textbf{q }$\displaystyle =\left(x^{0},\varphi\right)$};
\draw (460,302) node [anchor=north west][inner sep=0.75pt]   [align=left] {\textbf{p}$\displaystyle =(R,\varphi )$ \ \ \ \ \ \textbf{q }$\displaystyle =(r ,\overline{\varphi })$};

\end{tikzpicture}
	\caption{Cartesian, prolate spheroidal and Gaussian normal coordinate systems represented on the plane with constant $\theta$ for the points ${\bf q}$ and ${\bf p}$. The point ${\bf p}$ lies on a prolate ellipsoid and is the nearest point on this surface to the point ${\bf q}$.}
\label{EllipsoidThetaCte}
\end{center}
\end{figure}

The normal vector ${\bf n}=(n_x,n_y)$ can be computed as the following. It is orthogonal to the tangent vector $\partial{\bf p}/\partial\varphi$ to ${\bf p}$. Using the above equation, we can compute that $\partial{\bf p}/\partial\varphi =(Ra\cos\varphi,-Rb\sin\varphi)$. From the relation ${\bf n}\cdot\partial{\bf p}/\partial\varphi=0$, we obtain that
\begin{equation}
{\bf n}
=
\frac{(b\sin\varphi,a\cos\varphi)}{\sqrt{a^2\cos^2\varphi+b^2\sin^2\varphi \ }}
=
\frac{(b^2x_p,a^2z_p)}{\sqrt{a^4z_p^2+b^4x_p^2 \ }}
\end{equation}
Expressing ${\bf p}$, ${\bf q}$ and ${\bf n}$ in Cartesian coordinates, the relation~(\ref{pandq}) becomes
\begin{equation}
(x,z)=(x_p,z_p)+x^0\frac{(b^2x_p,a^2z_p)}{\sqrt{a^4z_p^2+b^4x_p^2 \ }}
\end{equation} 
Using relations~(\ref{coordinates}), we obtain
\begin{equation}
\label{E5}
\hspace{-0.5cm}
(ra\sin\overline{\varphi},
rb\cos\overline{\varphi})
=
(Ra\sin\varphi,Rb\cos\varphi)
+
x^0\frac{(b\sin\varphi,a\cos\varphi)}{\sqrt{a^2\cos^2\varphi+b^2\sin^2\varphi}}
\end{equation} 
We divide the first component of the above vector by $a$ and the second one by $b$. Summing up the squared of these two terms yields to 
\begin{eqnarray}
\label{E6}
r^2=
R^2+
\frac{2x^0R}{ab}
\sqrt{a^2\cos^2\varphi+b^2\sin^2\varphi \ }
+
\frac{(x^0)^2}{a^2b^2}
\frac{a^4\cos^2\varphi+b^4\sin^2\varphi}{a^2\cos^2\varphi+b^2\sin^2\varphi}
\end{eqnarray}
From the two above equations~(\ref{E5}) and (\ref{E6}), we obtain $r$ and $\overline{\varphi}$ from $x^0$ and $\varphi$. That is, once one knows the coordinates of a point in the Gaussian normal coordinate system, one can obtain its coordinates in the prolate spheroidal coordinate system. From equations~(\ref{coordinates}), one can then obtain the coordinates of this point in the Cartesian coordinate system.

Now, to obtain the coordinates of a point once one knows it in the Cartesian, or prolate spheroidal coordinate systems, it is necessary to invert these equations. From the Cartesian to the prolate spheroidal coordinate system, we have that:
\begin{eqnarray}
r=\sqrt{\frac{x^2}{a^2}+\frac{z^2}{b^2}},
\end{eqnarray}
and $\overline{\varphi}$ must satisfy
\begin{equation}
\cos\overline{\varphi}=
\frac{x}{ar}
\ \ \ \ \
{\rm and}
\ \ \ \ \
\sin\overline{\varphi}=
\frac{z}{br}.
\end{equation}

Now, to obtain the coordinates in the Gaussian normal coordinate system, we can proceed as the following. From equations~(\ref{E5}), we can reexpress
\begin{equation}
\label{E9}
x^0=\sqrt{a^2\cos^2\varphi+b^2\sin^2\varphi \ }
\frac{(ra\sin\overline{\varphi}-Ra\sin\varphi)}{b\sin\varphi}
\end{equation}
and
\begin{equation}
x^0=\sqrt{a^2\cos^2\varphi+b^2\sin^2\varphi \ }
\frac{(rb\cos\overline{\varphi}-Rb\cos\varphi)}{a\cos\varphi}
\end{equation}
Comparing these two equations we find that
\begin{equation}
\frac{b\cos\overline{\varphi}}{a\cos\varphi}+\frac{R}{r}
\left(
  \frac{a}{b}
  -
  \frac{b}{a}
\right)
=
\frac{a\sin\overline{\varphi}}{b\sin\varphi}
.
\end{equation}
By squaring this equation and replacing $\sin^2\varphi$ by $1-\cos^2\varphi$, we obtain $\cos\varphi$ as one of the roots of the fourth-degree polynomial:
\begin{eqnarray}
b^4\cos^2\overline{\varphi}
-\frac{2Rb^2}{r}(b^2-a^2)\cos\overline{\varphi}\cos\varphi
+
\nonumber
\\
\left(
\frac{R^2(b^2-a^2)^2}{r^2}
-
[
  b^4\cos^2\overline{\varphi}
  +a^4\sin^2\overline{\varphi}
]
\right)\cos^2\varphi
\label{4thPoly}
\\
\nonumber
+\frac{2Rb^2}{r}(b^2-a^2)\cos\overline{\varphi}\cos^3\varphi
-\frac{R^2(b^2-a^2)^2}{r^2}\cos^4{\varphi}
=0
\end{eqnarray}
The solution for $\cos\varphi$ as one of the roots of the above polynomial is a function of the ellipsoid parameters $a$, $b$ and $R$, and of the prolate spheroidal coordinates $\overline{r}$ and $\overline{\varphi}$.
For instance, for $r=1$, $a=1$, $b=1.5$, only two of the four roots are real numbers. One of them corresponds to the parameters of the closest point and the other to the furthest one. One can use equation~(\ref{E9}) to determine $x^0$ and choose the smallest value, and from this, determine the correspondent $\varphi$.

Note, however, that $x^0$ is the distance from a point to the ellipsoid. With this in mind, one can have a better visualization of the ellipsoid's properties in this coordinate system than one would obtain by looking at the method we above described.

\section{The uniform confinement potential}

A uniform confinement has a very simple expression in the Gaussian normal coordinate system, given by
\begin{equation}
\label{Vunif}
V=\frac{m\omega^2}{2}(x^0)^2
\end{equation}
which correspond to equation (\ref{Utrap}) for $\omega_2=0$.

However, it is not so simple to be expressed in the prolate spheroidal and in the Cartesian coordinate systems. In these coordinates, the variable $x^0$ is expressed as a function of $r$ and $\overline{\varphi}$, and of $x$ and $z$, respectively. However, to explicitly write this dependence, one must solve the fourth-degree polynomial~(\ref{4thPoly}) and follow the procedure explained in the previous section. The roots of a fourth-degree polynomial exist, their expressions are analytic and are easily solvable via numeric methods. However, the resulting expressions are lengthy and impractical for manual manipulation. Because of this complexity, in this paper, we restrict the explicit expression of the uniform confinement potential to the above expression~(\ref{Vunif}).  

\section{The bubble trap potential expressed in the Gaussian normal coordinate system} \label{AppBubbleTrap}

The confinement potential of a bubble trap is given by equations~(\ref{UbubbleCart}) and (\ref{UbubbleProlSphe}) in the Cartesian and prolate spheroidal coordinate systems, respectively, and, approximately, by equations (\ref {U72})--(\ref{U73}) in the Gaussian normal coordinate system.

This expression in Cartesian coordinates is obtained in literature, and the one in prolate spheroidal coordinate was obtained with algebraic manipulation in the main text. However, obtaining the bubble trap potential in the Gaussian normal coordinate system is more elaborated, being necessary to solve the fourth degree polynomial~(\ref{4thPoly}) for an analytic expression. Thus, we simply obtained an approximation of this potential in these coordinates. To do that, we perform a Taylor expansion of equations around $x^0=0$ on equation~(\ref{E6}):
\begin{eqnarray}
r\simeq & R+\frac{x^0}{ab}
\sqrt{a^2\cos^2\varphi+b^2\sin^2\varphi}
\\
\nonumber
&+\frac{(x^0)^2}{2Ra^2b^2}
  \left(
    \frac{a^4\sin^2\varphi+b^4\cos^2\varphi}{a^2\sin^2\varphi+b^2\cos^2\varphi}-(a^2\sin^2\varphi+b^2\cos^2\varphi)
  \right)+...
\end{eqnarray}
Then, we can approximate
\begin{equation}
	(r-R)^2\simeq
	\frac{\left(a^2\cos^2\varphi+b^2\sin^2\varphi\right)}{ab}(x^0)^2+...
\end{equation}
Inserting this expression into equation~(\ref{U72}), we obtain (\ref{U73})--(\ref{omegabbtrap}), which is the bubble trap potential expressed in the Gaussian normal coordinate system.

\

\

\end{document}